\documentclass[11pt]{article}

\usepackage[margin=1in]{geometry}

\usepackage{amsthm,amsmath,amsfonts,amssymb}
\usepackage[authoryear,round]{natbib}
\usepackage[colorlinks,citecolor=blue,urlcolor=blue]{hyperref}
\usepackage{graphicx}
\usepackage{bbm}
\usepackage{subfigure}
\usepackage{caption}
\usepackage{multibib}
\usepackage{algorithm}
\usepackage{algorithmic}
\usepackage[normalem]{ulem} 
\usepackage{xcolor}

\newcites{supp}{References}

\title{Topological inference on brain networks with application to lesion symptom mapping}

\author{
Yuan Wang\textsuperscript{1}\thanks{wang578@mailbox.sc.edu} \and
Jian Yin\textsuperscript{2}\and 
Nicholas Riccardi\textsuperscript{3}\and 
Dirk-Bart Den Ouden\textsuperscript{4}\and 
Julius Fridriksson\textsuperscript{5}\and 
Rutvik H. Desai\textsuperscript{6}
}

\date{}

\begin{document}

\maketitle

\begin{center}
\textsuperscript{1}Department of Epidemiology and Biostatistics, University of South Carolina, USA \\
\textsuperscript{2}Department of Biostatistics, City University of Hong Kong, China \\
\textsuperscript{3}Department of Communication Science and Disorders, University of South Carolina, USA \\
\textsuperscript{4}Department of Communication Science and Disorders, University of South Carolina, USA \\
\textsuperscript{5}Department of Communication Science and Disorders, University of South Carolina, USA \\
\textsuperscript{6}Department of Psychology, University of South Carolina, USA
\end{center}

\begin{abstract}
Persistent homology (PH) characterizes the shape of brain networks through persistence features. Group comparison of persistence features from brain networks can be challenging as they are inherently heterogeneous. A recent scale-space representation of persistence diagrams (PDs) through heat diffusion reparameterizes them using a finite number of Fourier coefficients with respect to the Laplace--Beltrami (LB) eigenfunction expansion of the domain, providing a powerful vectorized algebraic representation for group comparisons. In this study, we develop a transposition-based permutation test for comparing multiple groups of PDs using heat-diffusion estimates. We evaluate the empirical performance of the spectral transposition test in capturing within- and between-group similarity and dissimilarity under varying levels of topological noise and cycle location variability. In application, we propose a \emph{topological lesion symptom mapping} (TLSM) method based on the proposed framework. The method is applied to resting-state functional brain networks of individuals with post-stroke aphasia to identify characteristic cycles associated with varying levels of speech-language impairment.
\end{abstract}

\noindent \textbf{Keywords:} Topological data analysis; Persistent homology; Permutation test; Brain network; Lesion symptom mapping

\noindent \textbf{MSC Classification:} 97K80 (Primary), 92B15 (Secondary)

\section{Introduction}
\label{sec:intro}

Aphasia following stroke remains one of the most devastating causes of language impairment, disrupting communication and severely limiting recovery and quality of life \citep{Tsao2022}. Understanding how stroke lesions alter brain networks to produce specific language deficits is a central question in cognitive neuroscience. Traditional {\it lesion symptom mapping} (LSM) has been instrumental in identifying critical language regions, yet it largely treats brain areas as isolated units rather than components of an interconnected system \citep{Bates2003,Rorden2004}. However, because language and other cognitive abilities depend on distributed networks, follow‑up analyses that model how lesions disrupt region-to-region connectivity can complement traditional lesion-symptom mapping and provide a more complete picture of post‑stroke deficits. Recent studies of post-stroke aphasia have begun using connectome-based LSM (CLSM) to account for the connectivity patterns in brain connectomes/networks modeled as graphs \citep{Yourganov2016, Gleichgerrcht2017}. The approach still involves mass univariate testing, but on the connections between voxels or regions instead of the voxels or regions like voxel-based LSM (VLSM) or region-based LSM (RLSM).

The dimensional focus of traditional LSM approaches shifts from nodes in VLSM and RLSM to edges in CLSM. It fits how network topology is described with nodes, edges, and their higher-order combinations in simplicial homology - a branch of the mathematical field of topology \citep{Hatcher2001}. This prompts the question of whether an LSM framework can also be formulated on these higher-order topological structures. Moreover, existing LSM methods typically examine a single density threshold of the network, which provides only a static snapshot of connectivity. In contrast, cognitive functions such as language depend on network organization across multiple scales of connectivity. To capture this richer structure, we adopt a scale-free or all-scale approach using tools from computational topology. Specifically, we employ persistent homology (PH), which tracks how topological features such as clusters and loops (1-cycles) emerge and disappear as the network density changes. Each feature is represented by its “birth” and “death” values along this dynamic process, and together these form a persistence diagram (PD). The persistence of a feature reflects its relative importance or robustness within the network. In this study, we advance a {\em topological lesion symptom mapping} (TLSM) framework that accounts for the mesoscale homological features in two or more groups of brain networks with variable lesion presence. In TLSM, the node- and edge-level statistics used in standard LSM methods are extended to persistence features of cycles, which, depending on the application, can be generalized to higher-order topological structures such as polyhedra.

The innovation of TLSM does not, however, end with the dimensional and multiscale upgrade. We also innovate the way that permutation testing is conducted on persistence patterns of mesoscale homological features and extend it to multi-group settings. Current persistence descriptors are barcode and persistence diagram (PD), the original descriptors proposed by \citet{Edelsbrunner2002}, and persistence landscape (PL) \citep{Bubenik2015} and persistence image (PI) \citep{Adams2017}, both of which were developed when the demand increased for incorporating persistence features in statistical inference and machine learning models. A major challenge in statistical inference for PH arises because persistence features are inherently heterogeneous. Small perturbations in the underlying data—such as noise in fMRI signals or slight variations in network thresholding—can shift the birth and death times of topological features, causing points in PDs to move or even appear and disappear. This variability does not necessarily reflect biological differences but rather how homology computation reflects data fluctuations. As a result, persistence features from different subjects may have the same underlying topology but differ in their numerical representation, making direct parametric comparison unreliable based on stringent distributional assumptions that are rarely met in practice. To address this, we adopt permutation testing, a nonparametric approach that makes no assumptions about feature distributions and is therefore well suited for complex, non-Euclidean data structures such as PDs. A few approaches have been developed to overcome the computational bottleneck for permutation testing on persistence features. The {\em exact topological inference} approach allows for fast permutation of monotone functions built on birth or death times in barcodes with respect to the Komogorov-Smirnov (KS) distance \citep{Chung2019a}. This approach has quadratic run time that beats the exponential run time of standard permutation tests and has been extended to compare PLs \citep{Wang2019, Wang2021}. However, the approach is limited to comparing two features and not applicable for comparing between two sets of features. Another rapid permutation test based on transpositions does not require monotonicity and is applicable for comparing two sets of persistence features \citep{Chung2019b, Song2023}. It has allowed us to develop a unified framework for topological inference through heat kernel estimation of PDs.

Inference and learning approaches comparing PDs have been built on confidence band \citep{Fasy2014} to functional representations \citep{Chung2009, Pachauri2011, Bubenik2015, Reininghaus2015, Carriere2015, Chen2015, Adams2017}, as comparing raw PDs consisting of planar scatter points encoding birth and death times of topological structures often require point matching through, for instance, the Hungarian matching algorithm, which quickly becomes computationally prohibitive for large-scale data. It is also unclear how we may compare two sets of raw PDs. The functional representation approach overcomes the issue of the points on raw PDs having arbitrary locations and provides an effective framework for downstream comparison. In this approach, PDs essentially undergo a smoothing process, in some cases through a scale-space representation from kernels for heat diffusion of Dirac delta functions uniquely representing the points of PD \citep{Reininghaus2015}. However, existing kernel features on PD are typically convoluted, which lacks flexibility when performing resampling-based statistical inference procedures such as permutation testing. A new scale-space representation of PD was recently proposed based on the heat kernel (HK) estimation \citep{Kulkarni2020}, where the upper-triangular domain of PDs is represented using a finite number of Fourier coefficients with respect to the Laplace-Beltrami (LB) eigenfunction expansion of the domain. It provides a powerful vectorized algebraic representation for comparisons of PDs at the same coordinates, foregoing the need for matching across PDs due to their arbitrary point locations. Motivated by a topology-preserving spectral permutation test \citep{Wang2018}, we developed an inference procedure for comparing two sets of PDs estimated by the new scale-space representation by transposing the PD labels \citep{Wang2022}. By updating only the terms in an $L_2$-distance between the mean HK estimates of two sets of PDs involved in each transposition, computation becomes much faster than standard permutation testing that exchanges an arbitrary number of labels in each iteration. This inference procedure generalizes the method developed in \citep{Wang2018} for comparing persistence features of single-trial univariate signals, where the resampling takes place at the signal level and thus cannot be directly applied to images and networks. The inference framework now resamples at the feature level, which allows us to compare PDs of images and networks. We have also extended it to a new topological ANOVA (T-ANOVA) approach to compare across multiple groups of PDs without dimensionality reduction.

In summary, we establish a topological inference framework through a stable heat-kernel estimator of persistence diagrams (PDs). We developed two novel transposition-based tests under the framework: a two-sample test and multi-group topological ANOVA (T-ANOVA), and evaluated their empirical performance in simulation studies with heterogeneous topological noise and cycle location across multiple point clouds, as well as heterogeneous lesion presence in brain networks modeled from real data used in application. The proposed topological lesion symptom mapping (TLSM) method based on the topological inference framework is also the first of its kind. It was applied to compare resting-state functional brain networks of individuals with post-stroke aphasia, thus linking higher-order mesoscale features in functional brain connectivity with behavioral impairment of the disorder. We also empirically demonstrated how the TLSM approach complements traditional LSM methods relying on edge-wise correlation between disconnection strength and behavior.

\section{Preliminary}
\label{sec: preliminary}

This section provides background for our methodological development.

\subsection{Brain network filtration and persistence descriptors}

\begin{figure}[t!]
  \includegraphics[width = 1\linewidth]{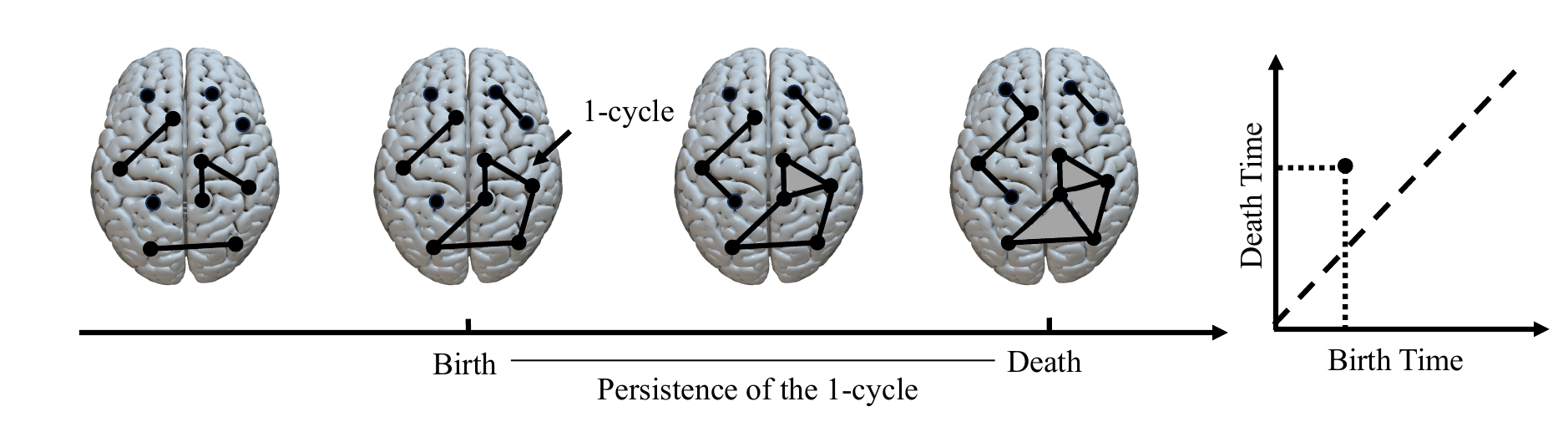}
  \caption{\label{fig: network_filtration} A 1-cycle emerges out of a Rips filtration constructed on a brain network and the corresponding persistence of the cycle is encoded in a persistence diagram (PD). }
\end{figure}

Brain networks are typically modeled as a weighted graph, with the edge weights given by a similarity, or dissimilarity, measure between the measurements on the nodes of the network \citep{Bassett2006,Bien2011}. Suppose we have a network represented by the weighted graph $G = (V, w)$ with the node set $V = \{1,\dots, p\}$ and unique positive undirected edge weights $w=(w_{ij})$ constructed from a similarity measure such as the absolute value of the Pearson's correlation between the blood oxygen level dependent (BOLD) signals of the $i$-th and $j$-th region of interest (ROI). We define the binary network $G_{\epsilon} = (V,w_{\epsilon})$ as a subgraph of $G$ consisting of the node set $V$ and the binary edge weights $w_{\epsilon}$ defined by
\begin{equation}
\label{eq: gf_threshold}
w_{ij,\epsilon} = \left\{ 
\begin{array}{cc}
1 & \text{if}~w_{ij}<\epsilon;\\
0 & \text{otherwise},
\end{array}\right.
\end{equation} 
where $w_{\epsilon}$ is the adjacency matrix of the subgraph $G_{\epsilon}$, which is a simplicial complex made up of nodes and edges. Any edge in $G$ with a weight $w_{ij}$ less than $\epsilon$ are connected. As we increase $\epsilon$, which we call the {\em filtration value}, more edges are included in the binary network $G_{\epsilon}$ and so the size of the edge set increases. Since edges connected in the network do not get disconnected again, we observe a sequence of nested subgraphs
\begin{equation}
\label{eq: rips_filtration}
G_{\epsilon_0}\subset G_{\epsilon_1} \subset G_{\epsilon_2} \subset \cdots \subset G, ~\text{for any}~\epsilon_0\le\epsilon_1\le\epsilon_2\le\cdots.
\end{equation}
This sequence of nested subgraphs make up a {\em Rips filtration} where two nodes with a weight $w_{ij}$ smaller than $\epsilon$ are connected, and the birth and death of {\em homological structures} in the form of clusters of nodes and cycles formed by more than 3 edges are tracked through the filtration \citep{Lee2011b,Lee2014}. We pair the birth and death times of clusters and cycles as the coordinates of scatter points on a planar graph $\{(a_i,b_i)\}_{i=1}^L$ in the {\em persistence diagram} (PD). The persistence of clusters and cycles is measured by the drop from their corresponding points to the $y = x$ line on the PD. Long persistence indicates that the corresponding cluster or cycle is more likely to be an underlying feature in the network. The dynamic changes in clusters in brain networks have been studied quite thoroughly in past studies \citep{Lee2011b}. As illustrated in Figure~\ref{fig: network_filtration}, a 1-cycle emerges out of a Rips filtration constructed on a brain network and the corresponding persistence of the cycle is encoded in the PD. Here we focus on one-dimensional cycles or 1-cycles for the purpose of lesion symptom mapping on resting-state functional brain networks. A 1-cycle in this context corresponds to a closed loop of connections among multiple regions, reflecting redundant or alternative communication pathways that support integrative processing. Focusing on 1-cycles enables characterization of how stroke lesions disrupt these mesoscale connectivity loops, particularly across frontal, temporal, and parietal regions involved in language. A reduction or disappearance of such loops may indicate a loss of compensatory connectivity, providing a biologically interpretable marker of functional disconnection in aphasia.

Note that there is a popular alternative approach to constructing filtration on brain networks, namely {\em graph filtration}. An array of powerful topological inference and learning methods for brain networks have been advanced on the basis of graph filtration \citep{Chung2015b,Chung2019a,Chung2019b,Song2023}. It differs from the Rips filtration in that triangles also count as 1-cycles and there are no higher-dimensional cycles beyond 0- and 1-cycles. Although we do not explore higher-dimensional cycles for lesion symptom mapping in this study, we would like to have the option in future studies and so set up our topological inference framework based on the Rips filtration.

\subsection{Heat kernel representation of persistence diagram}
\label{sec: hk}

Since PDs do not form a vector space, they do not possess a natural statistical framework \citep{Bubenik2015} and requires additional manipulation such as kernel smoothing for downstream statistical analysis. As with all noisy data, smoothing is needed for reducing noise (typically random, often artifactual) to better reveal the underlying data structure. We could either smooth data used to construct the networks or smooth persistence  descriptors such as PD. In principle, topological noise and artifacts should be better removed with the latter approach as persistence descriptors are designed to capture topological structures, be they inherent or transient. Another important reason for smoothing PDs is that the heterogenous nature of raw PDs makes it difficult to perform various algebraic operations for statistical inference. Various smoothing methods have been applied to PDs such that statistical inference can be directly performed on them. Beginning with the work of \citet{Chung2009}, each PD is discretized using the the uniform square grid and a concentration map is then obtained by counting the number of points in each pixel, which is equivalent to smoothing PD with a uniform kernel. This approach is somewhat similar to the voxel-based morphometry \citep{Ashburner2000}, where brain tissue density maps are used as a shapeless metric for characterizing concentration of the amount of tissue. \citet{Pachauri2011} followed up the approach by smoothing the PD by a Gaussian kernel centered at every point. Later, \citet{Bubenik2015} proposed the persistence descriptor PL by representing the PD as a function in the Banach space $L_p(\mathbb{R}^2)$ aimed at statistical analysis. PL is easily invertible to a PD, but overemphasizes the high-persistence features. To account for the overall pattern of persistence features, a persistence scale-space (PSS) kernel approach was then proposed by \citet{Reininghaus2015}, where the points in PD are treated as heat sources modeled as Dirac-delta functions and used as an initial condition for a heat diffusion problem with a Dirichlet boundary condition on the diagonal. The closed-form solution of the diffusion problem is an $L_2(\Omega)$ function obtained by convolving the initial condition with a Gaussian kernel, with $\Omega = \{(x,y)\in\mathbb{R}^2: y\ge x\}$ being the closed half plane above the diagonal line $y = x$, and the feature map from the PDs to $L_2(\Omega)$ at a fixed scale yields the PSS kernel. The Hilbert space structure of $L_2(\mathbb{R}^2)$ can be used to construct a PL kernel similar to PSS \citep{Reininghaus2015}. The relatively new persistence descriptor PI sampled at discrete uniform grid to produce homogenous vectorized data out of PDs \citep{Adams2017}. PIs live in Euclidean space and are therefore amenable to a broader range of learning techniques than PLs \citep{Adams2017}. A new heat kernel representation for PDs has recently been proposed by \citet{Kulkarni2020}, which not only allows a non-convoluted vectorized representation for comparisons at the same coordinates of PDs but also smoothing PD at different scales. It has also been combined with transposition test, a novel permutation testing approach, for fast inference on PDs \citep{Wang2022}.

\section{Methods}
\label{sec: methods}

Heat kernel (HK) representation has been established as a smoothing framework for noisy measurements on a general manifold $\mathcal{M}\subset\mathbb{R}^d$ \citep{Chung2007, Chung2014.MICCAI} (background details in Appendix). To construct a HK representation of PD, we restrict the domain of diffusion to $\mathcal{M} = \mathcal{T} = \{(x,y)\in\mathbb{R}^2: y>x\}$, i.e. the upper triangular region above the diagonal line $y=x$ where the scatter points of the PD $D = \{(a_i,b_i)\}_{i=1}^P$ are located. We constrain $\mathcal{T}$ within a certain range, such as standardizing the coordinates of the PD, so that $\mathcal{T}$ is bounded. Consider heat diffusion equation
\begin{equation}
\label{eq: heat}
\frac{\partial h(\sigma, p)}{\partial \sigma} = \Delta h(\sigma, p)
\end{equation}
with the initial condition
$\quad h(\sigma=0,p) = \sum_{i=1}^P \delta_{(a_i, b_i)} (p),$
where $\delta_{(a_i, b_i)}$ is the Dirac-delta function at $(a_i, b_i)$. The scatter points in the PD serve as the heat sources of the diffusion process. To simplify notation, we will refer to any series $h(\sigma,p)$ as $h_{\sigma}(p)$ as the bandwidth $\sigma$ is fixed. A unique solution to \eqref{eq: heat} is given by the HK expansion
\begin{eqnarray}
\label{eq: wfs}
h_{\sigma}(p) &=& \int_{\mathcal{T}} K_{\sigma}(p,q) h_0(q) \;d\mu(q) \nonumber \\
& = &\sum_{k = 0}^{\infty}e^{-\lambda_k\sigma}f_k\psi_k(p),
\end{eqnarray}
where 
\begin{equation}
\label{eq: hk_pd}
K_{\sigma}(p,q) = \sum_{k = 0}^{\infty}e^{-\lambda_k\sigma}\psi_k(p)\psi_k(q), p, q\in\mathcal{T},
\end{equation} 
is the HK with respect to the eigenfunctions $\psi_k$ of the LB operator $\Delta$ satisfying
$\Delta \psi_k(p) = \lambda_k \psi_k(p)$
for $p \in \mathcal{T}$,
and
\begin{equation} 
\label{eq: fourier}
f_k = \big\langle h_0,\psi_k\big\rangle=\int_{\mathcal{T}}  h_0(q)  \psi_k(q) \; d\mu(q)\\
= \sum_{i=1}^P \psi_k(a_i,b_i) 
\end{equation}
are the Fourier coefficients with respect to the the LB eigenfunctions. The first eigenvalue $\lambda_0 =0$ of the LB operator corresponds to eigenfunction $ \psi_0 = \frac{1}{\sqrt{\mu(\mathcal{T})}}$, where $\mu(\mathcal{T})$ is the area of the triangular region $\mathcal{T}$ and $\sigma$ is the bandwidth of the HK.

\begin{figure}[b!]
  \includegraphics[width = 1\linewidth]{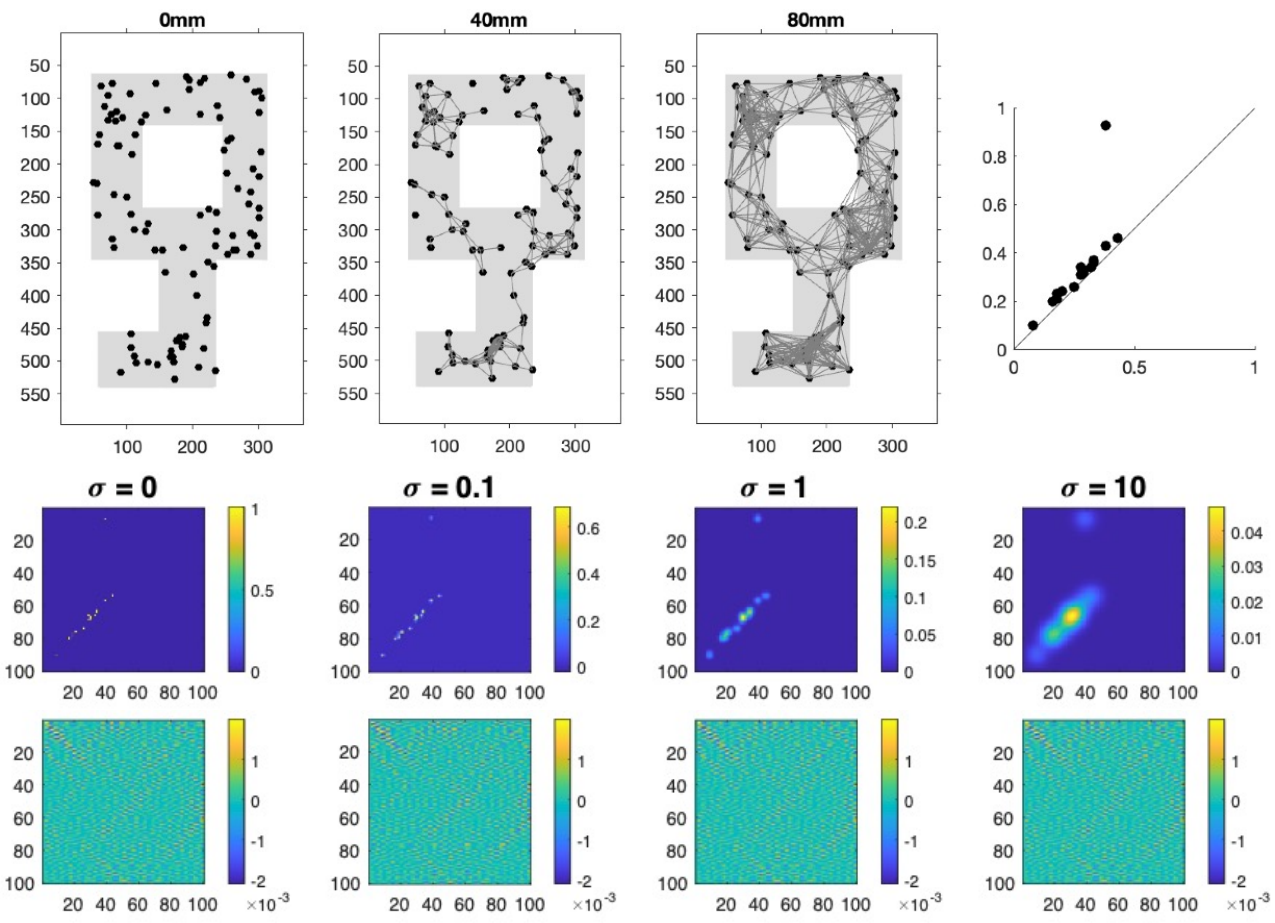}
  \caption{\label{fig: keyhole_pd_bandwidths} Top half: Left Three: The evolving 1-skeleton of a 100-point point cloud sampled from a key shape with a cycle. Right: PD from the Rips filtration constructed on the 1-skeletons of the point cloud. The point in the PD that corresponds to the key cycle stands out with high persistence - much further away from the diagonal ($y = x$) line than the rest of the points. Bottom half: Heat kernel (HK) smoothing of the PD from the Rips filtration through Laplace-Beltrami (LB) eigenfunctions with respect to the bandwidths $\sigma = 0$ (original PD), 0.1, 1, 10. Top Row: Smoothed PDs. Bottom Row: Corresponding Fourier coefficients with respect to the LB eigenfunctions presented in matrix form.} 
\end{figure}

The HK expansion \eqref{eq: wfs} provides a vectorized representation of the PD $D$ so that we can compare across PDs at the same coordinates while not having to recompute the coefficients $f_k$ for HK-represented PDs at different scales. In practice, we include sufficiently large $\kappa$ terms to approximate the HK expansion:
\begin{equation}
\label{eq: wfs_finite}
h^{\kappa}_\sigma(p) = \sum_{k = 0}^{\kappa}e^{-\lambda_k\sigma}f_k\psi_k(p),
\end{equation}
which we refer to as the degree-$\kappa$ HK estimate of the given PD. When $\sigma = 0$, we can completely recover the initial scatter points. As $\sigma \to \infty$, it is essentially smoothing the PD with a uniform kernel on $\mathcal{T}$. Figure~\ref{fig: keyhole_pd_bandwidths} shows how a point that corresponds to a 1-cycle stands out with high persistence in the PD from the Rips filtration constructed on a 100-point point cloud sampled from a key shape with a hole, as well as the HK smoothing of a PD with respect to the bandwidths $\sigma = 0, 0.1, 1, 10$. Note that the Fourier coefficients $f_k$ remain the same for all $k$ when constructing the HK expansion at different diffusion scale $\sigma$. 

As a distance measure for the HK-estimated PDs, we use the $L_2$-distance between the functions $h^1, h^2 \in L_2(\mathcal{T})$ defined as
\begin{equation}
\label{eq: l2}
\| h^1 - h^2\|^2_2 = \sum_{k=0}^{\infty}e^{-\lambda_k\sigma}(h^1_{k} - h^2_{k})^2,
\end{equation}  
where the $h^1_{k}$ and $h^2_{k}, k = 0, \dots, \infty,$ are the respective Fourier coefficients of $h^1$ and $h^2$ as defined in \eqref{eq: fourier} with respect to the LB eigenfunctions.

In the standard kernel setup, we have the feature map 
$$\Phi_{\sigma}: \mathcal{D}\rightarrow L_2(\mathcal{T}),$$
where $L_2(\mathcal{T})$ is the space of square integrable functions on $\mathcal{T}$ with the $L_2$-distance between the functions $g^1, g^2 \in L_2(\mathcal{T})$ defined as
\begin{equation}
\| g^1 - g^2\|^2_2 = \sum_{k=0}^{\infty}e^{-\lambda_k\sigma}(g^1_{k} - g^2_{k})^2,
\end{equation}  
where the $g^1_{k}$ and $g^2_{k}, k = 0, \dots, \infty,$ are the respective Fourier coefficients of $g^1$ and $g^2$ as defined in \eqref{eq: fourier} with respect to the LB eigenfunctions $\psi_k, k = 0,\dots,\infty$. Given bandwidth $\sigma > 0$, 
$$\Phi_{\sigma}(D) = h_{\sigma} = \sum_{k = 0}^{\infty}e^{-\lambda_k\sigma}f_k\psi_k(p), p\in D,$$ 
as defined in \eqref{eq: wfs} for a PD $D\in\mathcal{D}$. This feature map corresponds to the kernel 
$$K_{\sigma}(D_1,D_2)=\langle \Phi_{\sigma}(D_1),\Phi_{\sigma}(D_2)\rangle_{L_2(\mathcal{T})},$$ 
an explicit form of which is given by \eqref{eq: hk_pd}:
\begin{equation}
K_{\sigma}(p,q) = \sum_{k = 0}^{\infty}e^{-\lambda_k\sigma}\psi_k(p)\psi_k(q), p\in D_1, q\in D_2.
\end{equation}
We can show stability of the heat kernel
\begin{equation}
\|K_{\sigma} * g^1 - K_{\sigma} * g^2 \|_2   \leq \|g^1 - g^2\|_2
\end{equation}
as follows: The integral version of Jensen's inequality is
$$\phi \left(\int w(x) \;dx \right) \leq \int \phi ( w(x)) \;dx$$
for convex function $\phi$ \citep{Matkowski1994}. Following Jensen's inequality, 
\begin{eqnarray}
\|K_{\sigma} *g(p)\|_2^2  &=& \int_{\mathcal{T}} \left\vert\int_{\mathcal{T}} K_{\sigma}(p,q) g(q) \; d \mu(q)  \right\vert^2  \;d \mu(p) \\
     &\leq &         \int_{\mathcal{T}} \int_{\mathcal{T}} K_{\sigma}(p,q) \| g(q) \|^2 \; d \mu(q)   \;d \mu(p) \\
     &= & \int_{\mathcal{T}}  \left\vert g(q) \right\vert^2 \int_{\mathcal{T}} K_{\sigma}(p,q)  \; d \mu(p)   \;d \mu(q) \\
     &=&  \int_{\mathcal{T}}  \left\vert g(q) \right\vert^2  \;d \mu(q). 
\end{eqnarray}
We used the fact heat kernel $K_{\sigma}(p,q)$ is a probability distribution such that
$$\int_{\mathcal{T}} K_{\sigma}(p,q)  \; d \mu(p)=1.$$
Hence
$$\| K_{\sigma} *g(p)   \|_2  \leq  \|  g(p) \|_2$$
showing HK smoothing on PD is a contraction map \citep{Chung2018.EMBC}.
Letting $g=g^1-g^2$, we have the stability results. The HK smoothing reduces the topological variability in PD.

\begin{figure}[b!]
  \includegraphics[width = 1\linewidth]{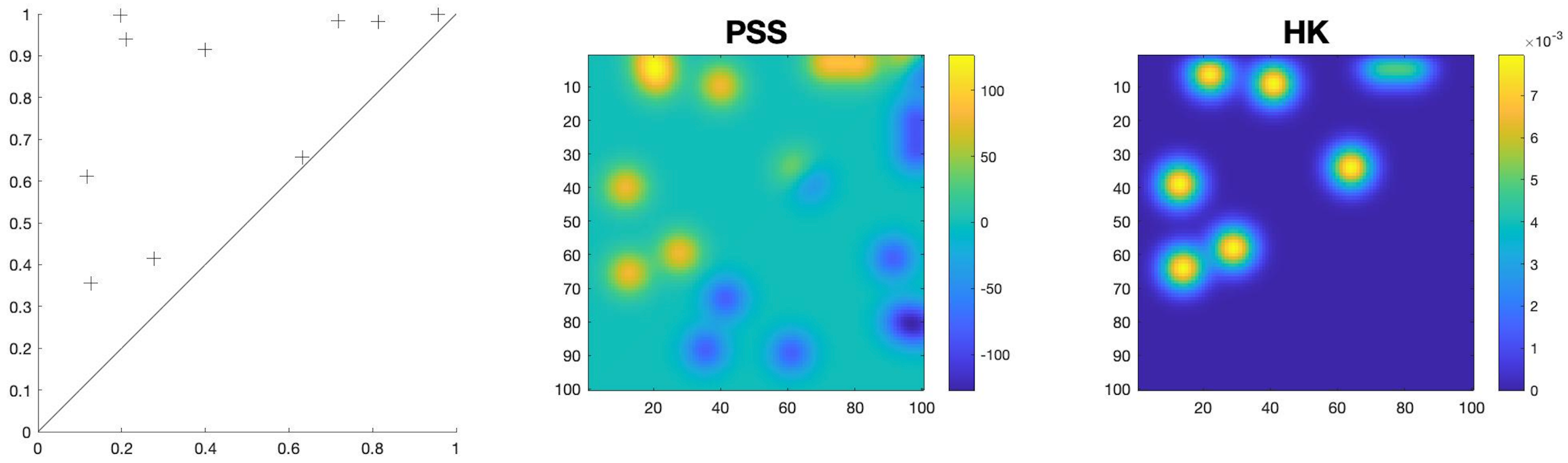}
  \caption{\label{fig: pd_pss_hk}An example of a PD and PSS- and HK-estimated versions.}
\end{figure}

We use a simple example with each of two PDs containing one of the two points $(-\lambda,\lambda)$ and $(-\lambda + 1,\lambda + 1)$ \citep{Reininghaus2015}, as an illustration of the stability of the kernel smoothing procedures. When comparing two PDs, the $L_2$-distance induced by the HK does not overweight any points in the PDs, as the distance between the two points is $\sum_{k=0}^{\infty}e^{-\lambda_k\sigma}(\psi_k(-\lambda,\lambda)-\psi_k(-\lambda+1,\lambda+1))^2$, which remains constant as $\lambda\rightarrow\infty$. In contrast, the PL-induced kernel distance is dominated by variations in the points of high persistence in the PDs, as the distance between the two points grows in the order of $\sqrt{\lambda}$ and is unbounded, whereas the Wasserstein distance and PSS-induced kernel distance do not overemphasize the high-persistence points, as the distance between the two points asymptotically approach a constant as $\lambda\rightarrow\infty$ \citep{Reininghaus2015}. While the PSS kernel representation, like our HK representation of PD, also uses an exact solution to the heat diffusion problem with the original PD as the initial condition (Figure~\ref{fig: pd_pss_hk}), the implicit form of the solution is difficult to manipulate for cost-effective resampling-based statistical inference. It is likewise difficult to manipulate the Wasserstein distance and PL-induced distance for the same purpose.

\subsection{Permutation test on HK-estimated PDs}
\label{sec: transposition}

Existing kernel features on PD have been shown theoretically and empirically to work well with machine learning frameworks \citep{Reininghaus2015, Adams2017} but are typically convoluted, which lacks flexibility when performing resampling-based statistical inference procedures such as permutation testing. Our past studies have shown powerful applications of the series representation of the heat diffusion problem, such as comparing the persistence features of brain signals through built permutation test based on HK estimates of signals \citep{Wang2018}, where we studied how topology of signals is preserved by permuting Fourier coefficients of sine and cosine basis functions. The approach provides a ground for permutation testing based on spectral components. The downside, however, is the computational load, with spectral permutation of single-trial signals requiring hours on end to converge. 

Here, we use the HK for PD smoothing and subsequent statistical inference based on the HK-estimated PDs. Once we have the HK estimates of PD, we can use them as the basis for permutation testing. But we do not use the standard approximate permutation test built on uniform sampling from the full set of permutations and typically applied in practice. The required number of permutations for convergence increases exponentially as the sample sizes increase. Even with sample sizes like $m=n=20$, the random permutation test requires significant computational resources if we compute the test statistic for each exchange of group labels. Instead, we adapt a {\em transposition}, which is defined as a permutation $\pi_{ij}$ that exchanges the $i$-th and $j$-th elements between ${\bf x}$ and ${\bf y}$ while keeping all the other elements fixed, i.e.
\begin{eqnarray*}
\pi_{ij}({\bf x}) & = & (x_1,\dots,x_{i-1},y_j,x_{i+1},\dots,x_m),\\
\pi_{ij}({\bf y}) & = & (y_1,\dots,y_{j-1},x_i,y_{j+1},\dots,y_n).
\end{eqnarray*}
Any permutation in $\mathbb{S}_{m+n}$ can be reached by a sequence of transpositions \citep{Chung2019b}. The random transposition is a random walk related to card shuffling problems and it is a special case of walk in symmetric groups \citep{Aldous1983,Aldous1986}. The walk between elements within ${\bf x}$ or ${\bf y}$ is also allowed but will not affect the computation a symmetric test functions. Instead of performing uniform random sampling in $\mathbb{S}_{m+n}$, we can perform a sequence of random walks and compute the test statistic at each walk. Consider walks in the two sample setting. We will determine how test statistic changes over each walk. Over random walk or transposition $\pi_{ij}$, the statistic changes from $L({\bf x}, {\bf y})$ to $L( \pi_{ij} ({\bf x}), \pi_{ij}({\bf y}))$. Instead of computing $L( \pi_{ij} ({\bf x}), \pi_{ij}({\bf y}))$ directly, we can compute it from $L({\bf x}, {\bf y})$ incrementally in {\em constant} run time by updating  the value of $L({\bf x}, {\bf y})$. If $L$ is an algebraic function that only involves addition, subtraction, multiplication, division, integer exponents, there must exist a function $M$ such that 
$ L(\pi_{ij}({\bf x}), \pi_{ij}({\bf y})) = M (L ({\bf x}, {\bf y}), x_i, y_i), \label{eq:increment}$
where the computational complexity of $M$ is constant \citep{Chung2019b}. For instance, basic test statistics such as the two-sample $t$-statistic and $F$-statistic are algebraic functions. If we take computation involving  fractional exponents as constant run time, then a much wider class of statistics such as correlations can all have iterative formulation  with constant run time. In the case of computing two-sample $t$-statistic with $m$ and $n$ samples directly, we need to compute the sample means, which takes $O(m)$ and $O(n)$ algebraic operations each. Then need to compute the sample variances and pool them together, which requires $O(3m+2)$ and $O(3n+2)$ operations each. Combining the numerator and denominator in $t$-statistic takes $O(16)$ operations. Thus, it takes total $O(4(m+n)+20)$ operations to compute the $t$-statistic at each permutation. In general, by only updating the terms in the test statistic affected by each transposition, the transposition test would require considerably less computational resources than the standard approximate permutation test. 

When we compare two groups of PDs with sample sizes $m$ and $n$, we assume under the null hypothesis that the functional means of the HK expansion of PDs are the same for both groups, for a fixed bandwidth $\sigma > 0$. The Fourier coefficients in the HK expansion of population PDs in the two groups are unknown. We estimate them with the HK expansion of sample PDs $\{f^{i}\}$ and $\{g^{j}\}$ from the groups approximated by their degree-$\kappa$ estimates: 
\begin{eqnarray}
\label{eq: wfs_pd}
f^i(p) &=& \sum_{k = 0}^{\kappa}e^{-\lambda_{k}\sigma}f^i_{k}\psi_k(p), i = 1,\dots, m,\\
g^j(p) &=& \sum_{k = 0}^{\kappa}e^{-\lambda_{k}\sigma}g^j_{k}\psi_k(p), j = 1,\dots, n,
\end{eqnarray}
where $f^i_{k}$ and $g^j_{k}$, $k = 0, \dots, \kappa$, are the Fourier coefficients with respect to the $k$-th LB eigenfunction $\boldsymbol{\psi}_k$. Their functional means are
\begin{eqnarray}
\label{eq: wfs_mean}
\bar{f}(p) &=& \sum_{k = 0}^{\kappa}e^{-\lambda_{k}\sigma}\bar{f}_{k}\psi_k(p), \\
\bar{g}(p) &=& \sum_{k = 0}^{\kappa}e^{-\lambda_{k}\sigma}\bar{g}_{k}\psi_k(p),
\end{eqnarray}
where $\bar{f}_k = \frac{1}{m}\sum_{i = 1}^m f^i_k$ and $\bar{g}_k = \frac{1}{n}\sum_{j = 1}^n g^j_k$ are the mean Fourier coefficients. We then use the $L_2$-norm difference $\left\|\bar{f} - \bar{g}\right\|^2_2$ between the functional means as a test statistic for measuring the group difference in HK expansion of the PDs. We can algebraically show that
\begin{equation}
\label{eq: wfs_l2}
\left\|\bar{f} - \bar{g}\right\|^2_2 = \sum_{k=0}^{\kappa} e^{-\lambda_k\sigma}(\bar{f}_k-\bar{g}_k)^2.
\end{equation}

In a standard approximate permutation test, the subject labels of the two groups are randomly exchanged. Here, we build the permutation test on transposition $\pi_{ij}$ that only exchanges the $i$-th and $j$-th subject labels between $\{f^i, i = 1,\dots, m\}$ and $\{g^j, j = 1,\dots,n\}$ and keeps all the other PDs fixed, i.e.
\begin{eqnarray}
\label{eq: trans}
\pi_{ij}(f^{1},\dots, f^{m}) & = & (f^{1},\dots, g^{j},\dots,f^{m}), \\
\pi_{ij}(g^{1},\dots, g^{n}) & = & (g^{1},\dots, f^{i}, \dots, g^{n}),
\end{eqnarray}
which we call a {\em spectral transposition}. Any permutation of the two groups of $m$ and $n$ subjects is reachable by a sequence of transpositions, which has been shown to be computationally much more efficient than the standard permutation testing procedure of exchanging all labels at once \citep{Chung2019b}. We generate the empirical distribution for the permutation test through the spectral transpositions. In one spectral transposition $\pi_{ij}$, we obtain the $L_2$-distance between the functional means of the degree-$\kappa$ HK estimates of PDs based on transposed labels:
\begin{equation}
\label{eq: l2_trans}
L_2(f,g) = \left\|\bar{f}' - \bar{g}'\right\|^2_2 = \sum_{k=0}^{\kappa} e^{-\lambda_k\sigma}(\bar{f}'_k-\bar{g}'_k)^2,
\end{equation}
where $$\bar{f}'_k=\bar{f}_k+\frac{1}{m}(g^j_k-f^i_k) \mbox{ and }  \bar{g}'_k=\bar{g}_k+\frac{1}{n}(f^i_k-g^j_k)$$ are the means of transposed Fourier coefficients. Since we know $\bar{f}_k$ and $\bar{g}_k$ already, we simply update the terms $\frac{1}{m}(g^j_k-f^i_k)$ and $\frac{1}{n}(f^i_k-g^j_k)$ affected by the transposition. The $p$-value of the spectral permutation test is then calculated as the proportion of $L_2$-distances in the empirical distribution exceeding the $L_2$-distance between the observed PDs. To ensure convergence, we perform upward of 100,000 transpositions in total until the $p$-value stabilizes. Labels are totally randomized after every sequence of a few hundred, e.g. 500, random transpositions to speed up the process.

\subsection{Topological analysis of variance via transpositions on HK-estimated PDs}

Topological analysis of variance allows us to assess within- and between-group similarity and dissimilarity in PDs across multiple groups. The challenge of applying an ANOVA procedure to raw PDs is that they do not have unique means \citep{Mileyko2011}. Thus, \citet{Heo2012} applied the standard ANOVA procedure to raw PDs reduced in dimensionality via Isomap. In contrast, our HK-estimates of PDs have well-defined functional means and $L_2$-distance through Fourier coefficients, which provides a natural framework for topological analysis of variance on PDs without any dimensionality reduction beforehand. 

To describe our heuristics in constructing an effective topological ANOVA framework, suppose the $K$ groups of HK-estimated PDs are expressed as follows:  
$$
\begin{array}{ccccc}
\text{Group 1}: & f^{11} & f^{12} & \cdots & f^{1n_1} \\
\text{Group 2}: & f^{21} & f^{22} & \cdots & f^{2n_2} \\
\vdots & \vdots \\
\text{Group K}: & f^{K1} & f^{K2} & \cdots & f^{Kn_K}
\end{array}
$$
Motivated by the standard ANOVA procedure, we could try and build an $F$-statistic comparing $K$ groups of HK-estimated PDs through the $L_2$-distance in \eqref{eq: wfs_l2}. A topological between-group sum of squares could take the form of 
\begin{equation}
\label{eq: tssb}
\sum_{i = 1}^{K}n_i||\bar{f}^i-\bar{f}||^2_2,
\end{equation}
and a topological within-group sum of squares the form of 
\begin{equation}
\label{eq: tssw}
\sum_{i = 1}^{K}\sum_{j = 1}^{n_i}||f^{ij} - \bar{f}^i||^2_2,
\end{equation}
where $f^{ij}$ is the HK-estimate of the $j$-th PD of the $i$-th group, $\bar{f}^i$ is the functional mean of the HK-estimates of PDs in the $i$-th group, and $\bar{f}$ is the grand functional mean over the HK-estimates of all PDs. The functional means would serve as the topological centroids. Ideally the $F$-statistic would follow $F$-distribution under some mild normality assumptions on the HK-estimated PDs, such as
$$
\frac{\sum_{i = 1}^{K}n_i||\bar{f}^i-\bar{f}||^2_2/K-1}{\sum_{i = 1}^{K}\sum_{j = 1}^{n_i}||f^{ij} - \bar{f}^i||^2_2/N-K}\sim F_{K-1,N-K},
$$
with $N = \sum_{i = 1}^{K}n_i$ and 
$$\sum_{i = 1}^{K}n_i||\bar{f}^i-\bar{f}||^2_2 \sim \chi^2_{K-1},$$ 
$$\sum_{i = 1}^{K}\sum_{j = 1}^{n_i}||f^{ij} - \bar{f}^i||^2_2 \sim \chi^2_{N-K}.$$
However, normality assumptions for heterogeneous features like PDs may be too strong to satisfy on multivariate data. 

Instead of fiddling with parametric constraints, we use a permutational ANOVA approach that bypasses the distributional issue and has found significant applications on multivariate data in response to complex experimental designs of ecological studies, where variables usually consist of counts of counts, percentage cover, frequencies, or biomass for a large number of species, and many other fields including chemistry, social sciences, agriculture, medicine, genetics, psychology, economics  \citep{Anderson2001, Anderson2017}. Here we build our test statistic for the permutational ANOVA based on pre-calculated pairwise distances between PDs so that no recalculation of distances is required after each transposition. We will only need to update the within- and between-group sums of distances after each transposition. We will refer to our topological ANOVA procedure as {\em T-ANOVA}, where we define the topological between-group sum of squares (TSSB) and topological within-group sum of squares (TSSW) based on sums of pairwise $L_2$-distances: 
\begin{eqnarray}
\label{eq: tss}
\text{TSSB} & = & \sum_{\substack{i,i'=1\\i<i'}}^{K}\sum_{j,j'} ||f^{ij}-f^{i'j'}||^2_2 \\
\text{TSSW} & = & \sum_{i=1}^K\sum_{j<j'}||f^{ij}-f^{ij'}||^2_2. 
\end{eqnarray}
We measure the between- and within-group disparity with the ratio statistic
\begin{equation}
\label{eq: ratio}
\phi = \frac{\text{TSSB}}{\text{TSSW}}.
\end{equation}
In each transposition, we randomly sample the group labels $i_1$ and $i_2$ out of the $K$ groups with respect to the proportions of the group sizes $n_i/N$. We then uniformly sample the subject labels $j_1$ and $j_2$ out of the $i_1$-th and $i_2$-th group respectively for transposition. We can prove by induction that any permutation between the groups can be reached by a sequence of transpositions through Theorem 1 in \citep{Chung2019b} showing any permutation between two groups can be reached by a sequence of transpositions.

In a transposition, we only update the pairwise $L_2$-distances in TSSB and TSSW affected by the transposition: 
\begin{eqnarray}
\text{TSSW}' 
& = & \sum_{i=1}^K\sum_{j<j'}||f^{ij}-f^{ij'}||^2_2\nonumber\\
& + & \sum_{j'\ne j_2}||f^{i_1j_1}-f^{i_2j'}||^2_2  - \sum_{j'\ne j_1}||f^{i_1j_1}-f^{i_1j'}||^2_2\label{eq: tssw_trans1}\\
& + & \sum_{j'\ne j_1}||f^{i_2j_2}-f^{i_1j'}||^2_2 - \sum_{j'\ne j_2}||f^{i_2j_2}-f^{i_2j'}||^2_2\label{eq: tssw_trans2}\\
& = & \text{TSSW}\nonumber\\
& + & \sum_{j'\ne j_2}||f^{i_1j_1}-f^{i_2j'}||^2_2  - \sum_{j'\ne j_1}||f^{i_1j_1}-f^{i_1j'}||^2_2\nonumber\\
& + & \sum_{j'\ne j_1}||f^{i_2j_2}-f^{i_1j'}||^2_2 - \sum_{j'\ne j_2}||f^{i_2j_2}-f^{i_2j'}||^2_2\nonumber,
\end{eqnarray}
where we adjust terms involving only groups $i_1$ and $i_2$ with \eqref{eq: tssw_trans1} and \eqref{eq: tssw_trans2}.
\begin{eqnarray}
\text{TSSB}' 
& = & \sum_{\substack{i,i'=1\\i<i'}}^{K}\sum_{j,j'} ||f^{ij}-f^{i'j'}||^2_2 \nonumber\\
& - & \sum_{j'\ne j_2}||f^{i_1j_1}-f^{i_2j'}||^2_2  + \sum_{j'\ne j_1}||f^{i_1j_1}-f^{i_1j'}||^2_2\label{eq: tssb_trans1}\\
& - & \sum_{j'\ne j_1}||f^{i_2j_2}-f^{i_1j'}||^2_2 + \sum_{j'\ne j_2}||f^{i_2j_2}-f^{i_2j'}||^2_2\label{eq: tssb_trans2}\\
& + & \mathbbm{1}(i'\ne i_1, i_2)\sum_{j'=1}^{n_{i'}}(||f^{i_2j_2}-f^{i'j'}||^2_2 - ||f^{i_1j_1}-f^{i'j'}||^2_2)\label{eq: tssb_trans3}\\
& + & \mathbbm{1}(i'\ne i_1, i_2)\sum_{j'=1}^{n_{i'}}(||f^{i_1j_1}-f^{i'j'}||^2_2 - ||f^{i_2j_2}-f^{i'j'}||^2_2)\label{eq: tssb_trans4}\\
& = & \text{TSSB}\nonumber\\
& - & \sum_{j'\ne j_2}||f^{i_1j_1}-f^{i_2j'}||^2_2  + \sum_{j'\ne j_1}||f^{i_1j_1}-f^{i_1j'}||^2_2\nonumber\\
& - & \sum_{j'\ne j_1}||f^{i_2j_2}-f^{i_1j'}||^2_2 + \sum_{j'\ne j_2}||f^{i_2j_2}-f^{i_2j'}||^2_2\nonumber,
\end{eqnarray}
where we adjust terms involving only groups $i_1$ and $i_2$ with \eqref{eq: tssb_trans1} and \eqref{eq: tssb_trans2}, terms involving groups other than $i_2$ that are affected by $i_1$ with \eqref{eq: tssb_trans3}, and terms involving groups other than $i_1$ that are affected by $i_2$ with \eqref{eq: tssb_trans4}. The ratio statistic is then updated to
\begin{equation}
\label{eq: ratio_trans}
\phi' = \frac{\text{TSSB}'}{\text{TSSW}'}.
\end{equation}

The $p$-value of the T-ANOVA test is then calculated as the proportion of $\phi'$ in the empirical distribution exceeding the $\phi$ between the observed PDs. We keep the transposed labels as the current labels on which we build the next transposition and randomize all labels every 500 transpositions to improve convergence rate.

\section{Performance Evaluation} 
\label{sec: simulations}

We conducted two sets of simulation studies to evaluate performance of the two-sample transposition test and T-ANOVA. All persistence patterns tested were of 1-cycles as they are more relevant to our application goal of understanding how circuits differ in the resting-state functional brain networks of individuals with post-stroke language deficits. 

\subsection{Simulation studies based on point clouds} In this section, we evaluated performance of the two-sample transposition test and T-ANOVA based on data simulated in point clouds.

\subsubsection{Performance of two-sample transposition test}

We investigated how the spectral transposition test detects underlying topological similarity and dissimilarity at the presence of topological noise and artifact. \\

\begin{figure}[b!]
\centering
  \includegraphics[width = 0.6\linewidth]{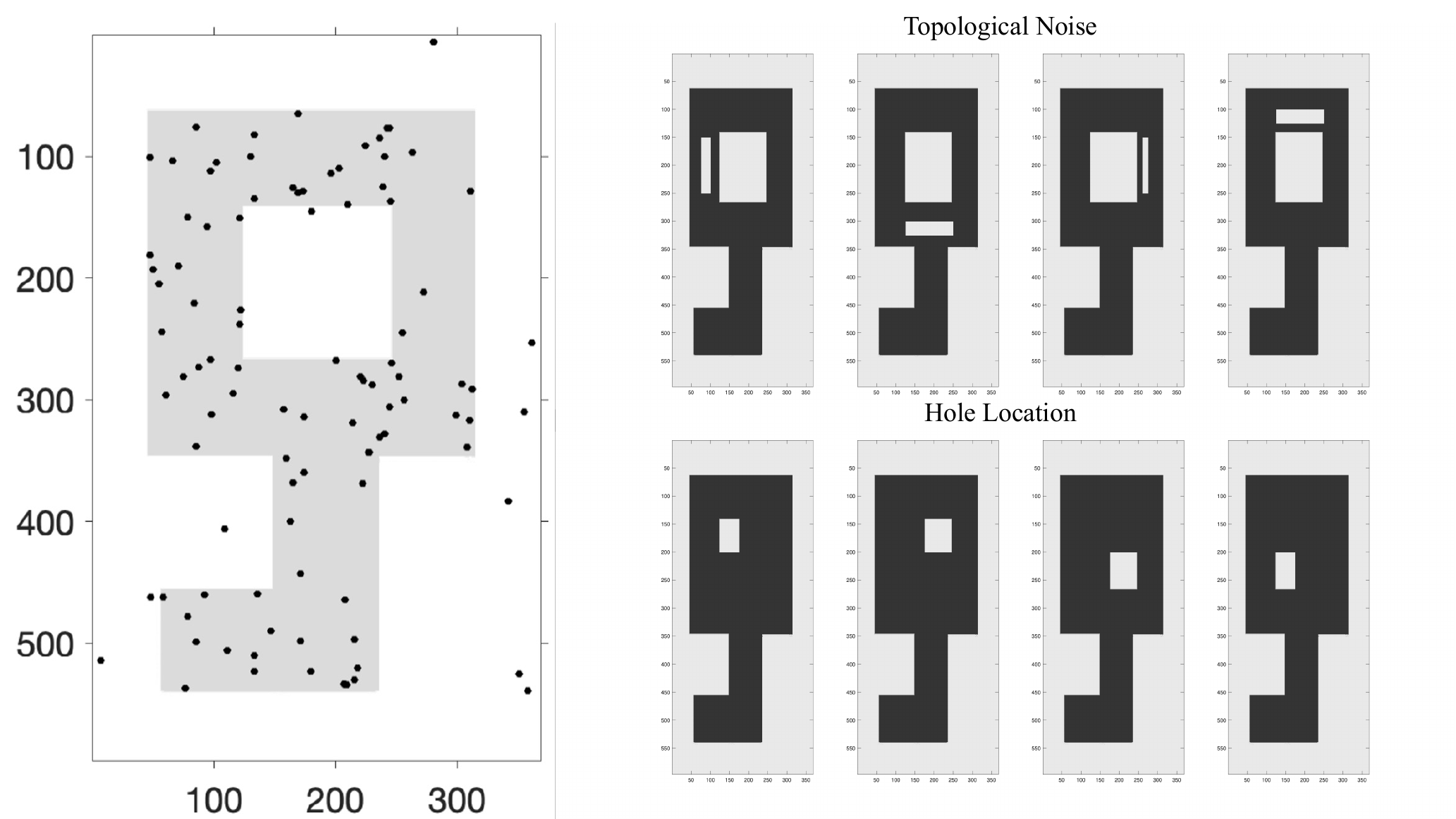}
  \caption{\label{fig: sim_key}Left: We randomly sampled 100 points from the image with an innate shape of a key. Top Right Row: Underlying key shape and possible locations of topological noise in the form of a small cycle. Bottom Right Row: Variants of the key shape in Group 2. They could appear in the 4 pre-specified forms or randomly out of the variants. }
\end{figure}

\noindent{\em Power of detecting cycle in structure.} We evaluated the power of the transposition test in detecting a key shape with a distinct hole/1-cycle (Figure~\ref{fig: sim_key} left), under different heterogeneity conditions. In each simulation, two groups of five 100-point point clouds were  generated: the 100 points in each point cloud of the first group were generated randomly from the part of the rectangular image, whereas the 100 points in each point cloud of the second group were generated randomly with a varied percentage (90\%, 95\%, 100\%) of points from the shape of the key. Rips filtration was constructed on each point cloud. The proposed spectral permutation test was then applied to compare the PDs of the Rips filtrations in the two groups. When there were respectively 90\%, 95\%, and 100\% points sampled from the shape of the key in the second group, the spectral permutation test rejected ($p$-value $<$ 0.05) the null hypothesis of no group difference in 91\%, 100\%, and 100\% of 100 simulations (corresponding means $\pm$ standard deviations of $p$-values: 0.0124$\pm$0.0327, 0.0041$\pm$0.0125, 0.0008$\pm$0.0057), showing that the test stays sensitive in detecting the group shape difference when points in the second group were not entirely sampled from the shape of the key. The experiments were repeated for 5 vs 5, 20 vs 20, and 100 vs 100 200-point point clouds. All showed 100\% of $p$-values $<$ 0.05 in 100 simulations (corresponding means $\pm$ standard deviations of $p$-values: 0.0033$\pm$0.0017, 0.0032$\pm$0.0017, 0.0034$\pm$0.0014 for 5 vs 5 when 90\%, 95\%, and 100\% points were respectively sampled from the shape of the key in the second group; $p$-values were all 0 in all the other cases).  \\

\noindent{\em Robustness of performance under variation of topological noise and cycle location.} \label{sec: sim_trans_robust} We conducted two studies to assess the robustness of the test when the underlying topological structure is "contaminated" with heterogeneous topological noise and when the underlying structure undergoes non-topological changes. We first evaluated the robustness of performance under heterogeneity of topological noise. In each of 100 simulations, we used the spectral transposition test to compare Group 1 of $m$ random samples with a varied percentage (90\%, 95\%, 100\%) of 100 or 200 points from the original key shape with Group 2 of $n$ random samples from the key shape 'contaminated' with topological noise in the form of a much smaller 1-cycle next to the keyhole with pre-specified (in such case $m = 4$ vs $n = 4$) or random locations (in such case $m = 5$ vs $n = 5$, $m = 20$ vs $n = 20$, or $m = 100$ vs $n = 100$). Figure~\ref{fig: sim_key} (top right row) shows the 4 possible locations of the topological noise in Group 2 and Table~\ref{tab: results_trans_robustness} shows the results for different percentage of points when the topological noise appears at pre-specified vs random locations. The spectral transposition test stayed robust to the topological noise in fixed and random locations. We then evaluated the robustness of performance under variation of cycle location. In each of 100 simulations, we used the spectral transposition test to compare Group 1 of $m$ random samples with a varied percentage (90\%, 95\%, 100\%) of 100 points from the original key shape with only the top left quarter of the keyhole left, with Group 2 of $n$ random samples with the same percentage of 100 points from the original key shape with a pre-specified (in such case $m = 4$ vs $n = 4$) or random (in such case $m = 5$ vs $n = 5$, $m = 20$ vs $n = 20$, or $m = 100$ vs $n = 100$) quarter of the keyhole left. Figure~\ref{fig: sim_key} (bottom right half) shows the 4 possible variants of the keyhole in Group 2 and Table~\ref{tab: results_trans_robustness} shows the results for different percentage of points when the variants appears at pre-specified vs random locations. The spectral transposition test stayed robust to the structural variants in fixed and random locations as these changes are not topological in nature. \\

 \noindent{\em Computational time.} The computational time of the spectral transposition test grew steadily as the group sample sizes grew. The mean time for each simulation run for $m = 4, 5$ vs $n = 4,5$ were between 7 and 10 seconds and standard deviation within 3 seconds. For $m = 20$ vs $n = 20$, the mean time for each simulation run were between 8 and 10 seconds and standard deviation within 3 seconds. For $m = 100$ vs $n = 100$, the mean time for each simulation run were between 9 and 11 seconds and standard deviation within 3 seconds.

\subsubsection{Performance of T-ANOVA}

In each of the simulation studies in this section, we tested the performance of the T-ANOVA in comparing three groups of point clouds simulated under different settings. The performance was compared against the standard PERMANOVA test \citep{Anderson2001}, as well as the topological analysis of variance test proposed by \citet{Heo2012} that runs the univariate ANOVA on dimensionality-reduced PDs by Isomap. \\

\begin{figure}
\centering
  \includegraphics[width = 0.6\linewidth]{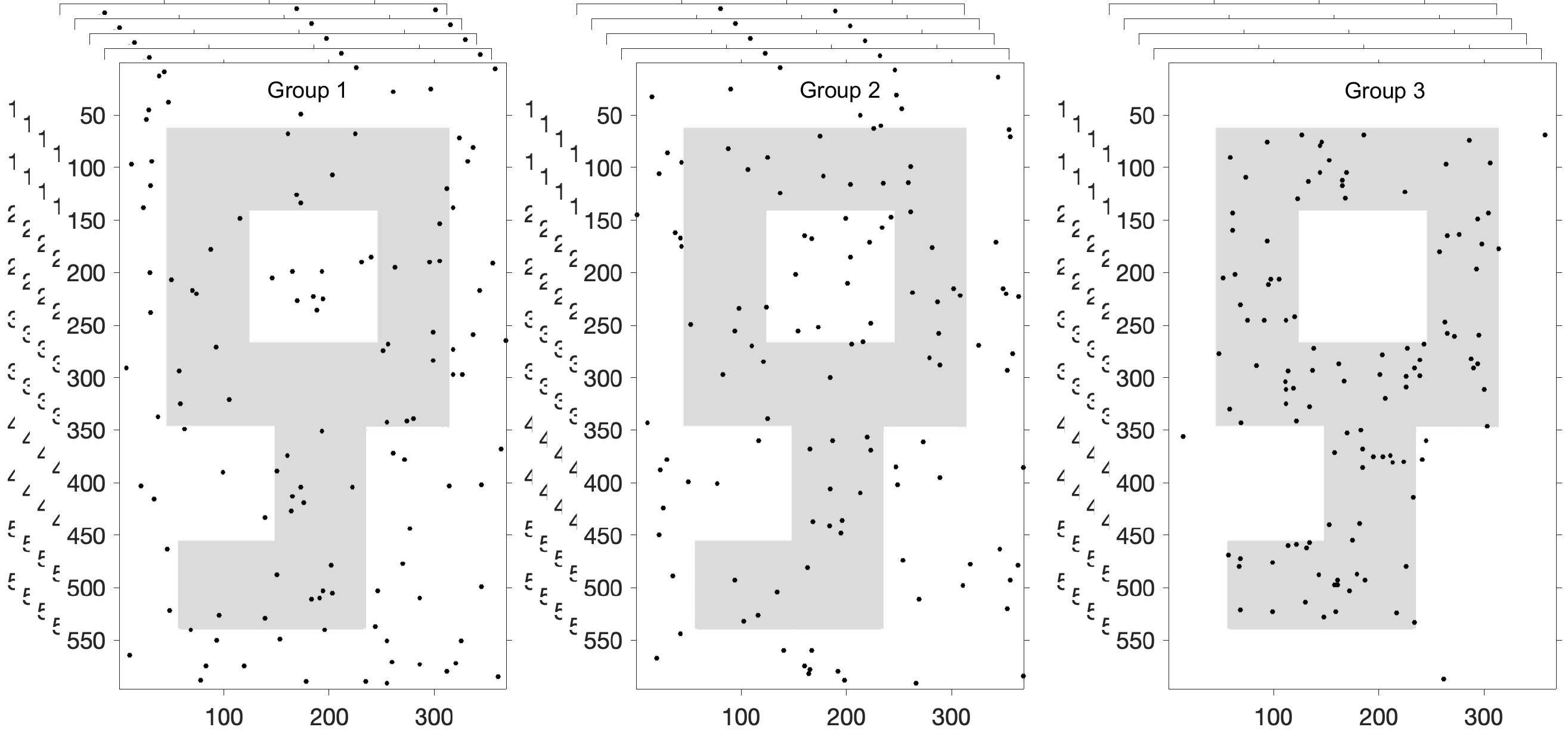}
  \caption{\label{fig: sim_anova_power}An example of $n_1=n_2=n_3=5$ 100-point point clouds where the 100 points in each point cloud of the first two groups were generated randomly from the part of the rectangular image, whereas the 100 points in each point cloud of the third group were generated randomly with 95\% of points from the shape of the key.}
\end{figure}

\noindent{\em Sensitivity in detecting differential cycle presence among multiple groups.} In each simulation, three groups of $n_1, n_2, n_3$ 100- or 200-point point clouds were generated, where the 100 or 200 points in each point cloud of the first two groups were generated randomly from the part of the rectangular image, whereas the 100 or 200 points in each point cloud of the third group were generated randomly with a varied percentage (90\%, 95\%, 100\%) of points from the shape of the key (Figure~\ref{fig: sim_anova_power}). Table~\ref{tab: results_anova_power} shows the results of the T-ANOVA test  in comparison with the other tests.\\

\noindent{\em Robustness under variation of noise and cycle location.} We conducted two studies to assess the robustness of the test when the underlying topological structure was "contaminated" with heterogeneous topological noise and when the cycle location shifts as in Section~\ref{sec: sim_trans_robust}. We first evaluated the robustness of T-ANOVA under heterogeneity of topological noise. In each of 100 simulations, Group 1, 2, 3 of respective $n_1, n_2, n_3$ random samples were generated with a pre-specified percentage (90\%, 95\%, 100\%) of 100 or 200 points from the original key shape "contaminated" with topological noise in the form of a much smaller 1-cycle next to the keyhole with pre-specified (in such case $n_1 = n_2 = n_3 = 4$) or random locations (in such case $n_1 = n_2 = n_3 = 5$, $n_1 = n_2 = n_3 = 20$, or $n_1 = 5, n_2 = 20, n_3 = 100$). The 4 possible locations of the topological noise in each group were the same as Figure~\ref{fig: sim_key}. We used the T-ANOVA test to compare PDs across the three groups. Table~\ref{tab: results_anova_robustness} (top half) shows the results of the T-ANOVA test in comparison with standard PERMANOVA and the topological ANOVA proposed by \citet{Heo2012}. We then evaluated the robustness of T-ANOVA under variation of cycle location. In each of 100 simulations, Group 1, 2, 3 of respective $n_1, n_2, n_3$ random samples were generated with a pre-specified percentage (90\%, 95\%, 100\%) of 100 or 200 points from the original key shape with a pre-specified (in such case $n_1 = n_2 = n_3 = 4$) or random (in such case $n_1 = n_2 = n_3 = 5$, $n_1 = n_2 = n_3 = 20$, or $n_1 = 5, n_2 = 20, n_3 = 100$) quarter of the keyhole left. The 4 possible variants of the key shape in each group were the same as Figure~\ref{fig: sim_key} . We used the T-ANOVA test to compare PDs across the three groups. Table~\ref{tab: results_anova_robustness} (bottom half) shows the results of the T-ANOVA test in comparison with the other tests. \\ 

\subsubsection{Summary of results} The methods were compared under these experimental settings for the first time. The results from both sets of experiments show that the performance of our T-ANOVA test was comparable with the two baseline methods in terms of robustness under variation of topological noise and cycle location, as well as sensitivity in detecting differential cycle presence among multiple groups. In comparison with PERMANOVA, the advantage of the transposition approach of T-ANOVA showed up in the steady growth of computational time as group sizes increased. Although Heo's ANOVA is comparable in performance as T-ANOVA, T-ANOVA does not require dimensionality reduction of PDs and can be flexibly applied when we want to focus on the persistence patterns of certain homological structures. As we see in the considerably more complicated network-based simulations below, Heo's ANOVA does not hold up the performance in some settings.

\subsection{Simulation studies based on real brain networks} \label{sec: sim_network} We also evaluated the performance of the proposed topological inference methods in comparing persistence patterns of groups of brain networks simulated as noisy variants of real resting-state functional brain networks described in Section~\ref{sec: data}. Simulation of cycles is not as straightforward as in the case of point clouds. Motivated by how stroke lesions affect the brain network topology, we designated various numbers of ROIs in a specific brain atlas as the nodes that we knocked out in a network. Any connections between the nodes and all the other nodes in the network were then removed, i.e. weights associated with the connections in the real resting-state functional connectivity matrix were set to 0, before any noise perturbations were added to the remaining weights in a simulation. Although this “node knockout” strategy is a simplification, it provides a clear and controlled way to mimic the functional disconnection that occurs when a lesion substantially overlaps with a cortical region. This binary manipulation allowed us to isolate topological effects of disconnection while avoiding the additional variability introduced by partially scaling edge weights, for which no physiologically validated model currently exists. We may explore more graded models of partial disconnection in future work. 

The studies were again grouped by whether the performance of two-sample test or T-ANOVA was under evaluation. Under each evaluation section, the following two settings were considered. 
\begin{itemize}
  \item Slight perturbations to network topology: Noisy perturbations were added to the same underlying real connectivity matrices in two or more groups. The groups thus generated are expected to be topologically similar and topological inference comparing the persistence patterns of 1-cycles through Rips filtration is expected to show a high average $p$-value on the null hypothesis of no topological difference in a number of simulations. 
  \item Differential lesion presence: Noisy perturbations were added to a group of real connectivity matrices and another or more group(s) of the same underlying connectivity matrices with various numbers of nodes and their connections to all other nodes removed to mimic lesion presence. The groups thus generated are expected to be topologically dissimilar and topological inference comparing the persistence patterns of 1-cycles through Rips filtration is expected to show a low average $p$-value on the null hypothesis of no topological difference in a number of simulations. 
\end{itemize}

\subsubsection{Performance of two-sample transposition test} In this section, two groups of networks were generated in each of 100 simulations. Different settings are as follows. Note that we used the two-sample version of Heo's ANOVA and PERMANOVA for comparison with the two-sample transposition test.\\
\noindent{\em Robustness under slight perturbations to network topology.} Firstly, 10, 20, or 50 resting-state functional connectivity matrices created from the JHU atlas of 189 ROIs were randomly selected out of the real dataset. In each of 100 simulations, two groups of independent Gaussian noise with mean 0 and standard deviation of 0.0001 were added to each of the edge weights in the connectivity matrices. \\
\noindent{\em Sensitivity in detecting differential lesion presence.} Two groups of independent Gaussian noise with mean 0 and standard deviation of 0.0001 were added to the same 10, 20, or 50 resting-state functional connectivity matrices per the JHU atlas randomly selected out of the real dataset. Then for the second group, 3 or 4 cluster(s) of 4 nodes in the frontal, temporoparietal, premotor, temporal lobes and their connections with all the other nodes were removed by setting the corresponding edge weights to 0. \\
\subsubsection{Performance of T-ANOVA} In this section, three groups of networks were generated in each of 100 simulations. Different settings are as follows.\\
\noindent{\em Robustness under slight perturbations to network topology.} In each of 100 simulations, three groups of independent Gaussian noise with mean 0 and standard deviation of 0.0001 were added to the same 10, 20, or 50 resting-state functional connectivity matrices per the JHU atlas randomly selected out of the real dataset. \\
\noindent{\em Sensitivity in detecting differential lesion presence.} Three groups of independent Gaussian noise with mean 0 and standard deviation of 0.0001 were added to the same 10, 20, or 50 resting-state functional connectivity matrices per the JHU atlas randomly selected out of the real dataset. For the second group, 1, 2, 3, or 4 cluster(s) of 4 nodes in the frontal, temporoparietal, premotor, temporal lobes and their connections with all the other nodes were removed by setting the corresponding edge weights to 0. \\

\begin{table}[h!]
\resizebox{\linewidth}{!}{
 \begin{tabular}{c|ccc}
\hline
   \multicolumn{4}{c}{\bf Performance of Two-sample Tests} \\ \hline
    \multicolumn{4}{c}{\bf Robustness under Slight Perturbations to Network Topology} \\ \hline
    Sample Size & Topological Transposition & Heo's ANOVA (Two sample) & PERMANOVA (Two sample) \\\hline
    10 vs 10 & $0.9973 \pm 0.0017$ & $0.9936 \pm 0.0050$ & $0.9970 \pm 0.0014$ \\
    20 vs 20 & $1.0000 \pm 0.0000$ & $0.9938 \pm 0.0047$ & $1.0000 \pm 0.0000$ \\
    50 vs 50 & $1.0000 \pm 0.0000$ & $0.9916 \pm 0.0063$ & $1.0000 \pm 0.0000$ \\ \hline
    \multicolumn{4}{c}{\bf Sensitivity in Detecting Differential Lesion Presence}\\\hline
    \multicolumn{4}{c}{ No Lesion Vs Frontal (4) \& Temporoparietal (4) \& Premotor (4)} \\ \hline
    Sample Size & Topological Transposition  & Heo's ANOVA (Two sample) & PERMANOVA (Two sample) \\\hline
    10 vs 10 & $0.0062 \pm 0.0020$ & $0.2845 \pm 0.2212$ & $0.0065 \pm 0.0004$ \\
    20 vs 20 & $0.0000 \pm 0.0000$ & $0.0101 \pm 0.0860$ & $0.0000 \pm 0.0000$ \\
    50 vs 50 & $0.0000 \pm 0.0000$ & $0.0049 \pm 0.0157$ & $0.0000 \pm 0.0000$ \\ \hline
    \multicolumn{4}{c}{ No Lesion Vs Frontal (4) \& Temporoparietal (4) \& Premotor (4) \& Temporal(4)} \\ \hline
    Sample Size & Topological Transposition & Heo's ANOVA (Two sample) & PERMANOVA (Two sample) \\\hline
    10 vs 10 & $0.0000 \pm 0.0000$ & $0.0006 \pm 0.0002$ & $0.0000 \pm 0.0000$ \\
    20 vs 20 & $0.0000 \pm 0.0000$ & $0.0000 \pm 0.0000$ & $0.0000 \pm 0.0000$ \\
    50 vs 50 & $0.0000 \pm 0.0000$ & $0.0000 \pm 0.0000$ & $0.0000 \pm 0.0000$ \\ \hline\hline
    \multicolumn{4}{c}{\bf Performance of ANOVA} \\ \hline
    \multicolumn{4}{c}{\bf Robustness under Slight Perturbations to Network Topology} \\ \hline
     \multicolumn{4}{c}{ No Lesion Vs No Lesion Vs No Lesion} \\ \hline
    Sample Size & T-ANOVA & Heo's ANOVA & PERMANOVA \\\hline
    10 vs 10 vs 10 & $1.0000 \pm 0.0000$ & $0.9999 \pm 0.0001$ & $1.0000 \pm 0.0000$ \\
    20 vs 20 vs 20 & $1.0000 \pm 0.0000$ & $0.9999 \pm 0.0001$ & $1.0000 \pm 0.0000$ \\
    50 vs 50 vs 50 & $1.0000 \pm 0.0000$ & $0.9998 \pm 0.0002$ & $1.0000 \pm 0.0000$ \\ \hline
    \multicolumn{4}{c}{\bf Sensitivity in Detecting Differential Lesion Presence}\\\hline
     \multicolumn{4}{c}{ No Lesion Vs No Lesion Vs Frontal (4) \& Temporoparietal (4) \& Premotor (4)} \\ \hline
    Sample Size & T-ANOVA  & Heo's ANOVA  & PERMANOVA  \\\hline
    10 vs 10 vs 10 & $0.0382 \pm 0.0091$ & $0.8798 \pm 0.2017$ & $0.0739 \pm 0.0027$ \\
    20 vs 20 vs 20 & $0.0002 \pm 0.0007$ & $0.0001 \pm 0.0000$ & $0.0024 \pm 0.0001$ \\
    50 vs 50 vs 50 & $0.0000 \pm 0.0000$ & $0.0197 \pm 0.1383$ & $0.0000 \pm 0.0000$ \\ \hline
     \multicolumn{4}{c}{ No Lesion Vs No Lesion Vs Frontal (4) \& Temporoparietal (4) \& Premotor (4) \& Temporal(4)} \\ \hline
   Sample Size & T-ANOVA  & Heo's ANOVA  & PERMANOVA  \\\hline
    10 vs 10 vs 10 & $0.0001 \pm 0.0003$ & $0.0007 \pm 0.0046$ & $0.0008 \pm 0.0001$ \\
    20 vs 20 vs 20 & $0.0000 \pm 0.0000$ & $0.0000 \pm 0.0000$ & $0.0000 \pm 0.0000$ \\
    50 vs 50 vs 50 & $0.0000 \pm 0.0000$ & $0.0000 \pm 0.0000$ & $0.0000 \pm 0.0000$ \\ \hline
\end{tabular}}
\caption{\label{tab: sim_network} Summary of mean$\pm$standard deviation of $p$-values of the two-sample tests or ANOVA in 100 simulations. Heo's ANOVA and PERMANOVA were adjusted under the two-sample test settings. Number in parentheses following a brain lobe means the number of nodes removed from that lobe, e.g. frontal (4) means 4 nodes in the frontal lobe and their connections to all the other nodes in the network were removed.}
\end{table}

\begin{figure}[h!]
\centering
  \includegraphics[width = 0.5\linewidth]{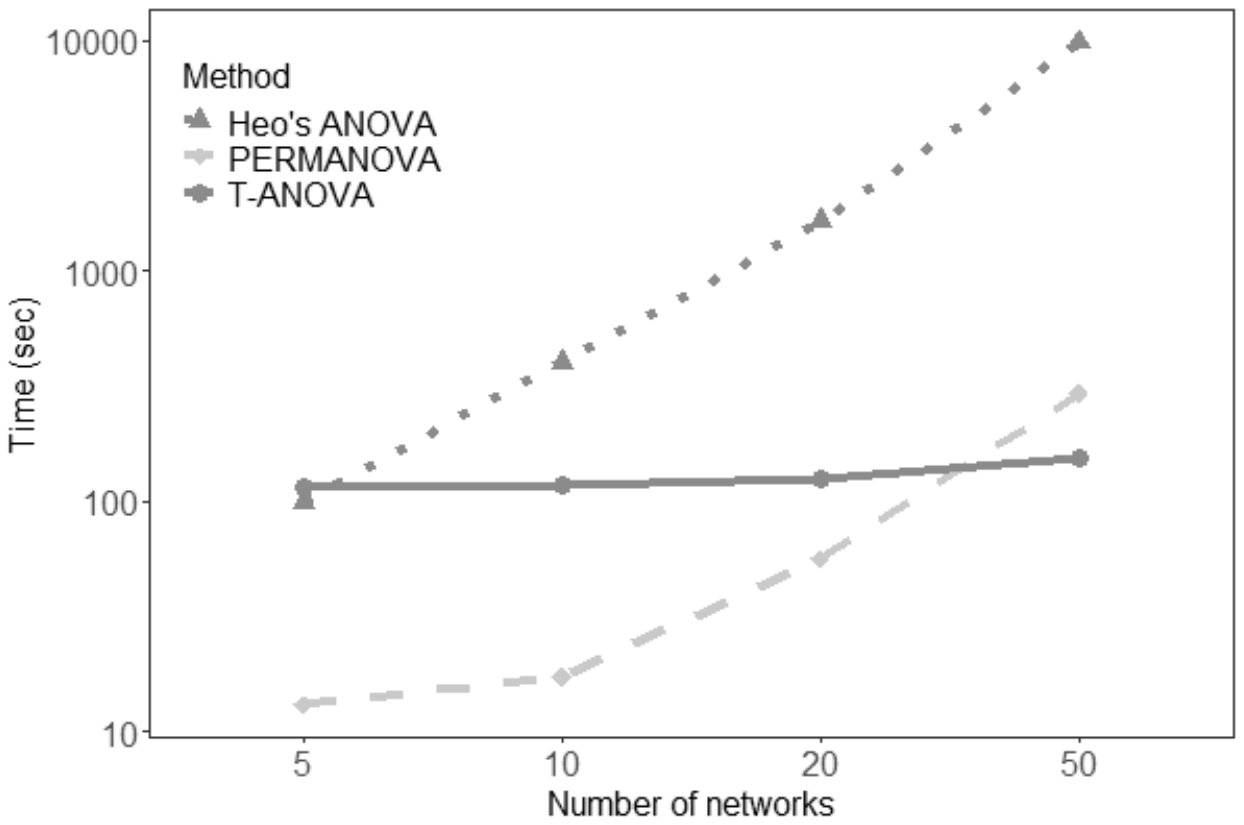}
  \caption{\label{fig: sim_time} Computational times of the three ANOVA methods. The time axis has been log-scaled.}
\end{figure}

\subsubsection{Summary of results} The results from both sets of experiments are summarized in Table~\ref{tab: sim_network} that shows that all methods maintain excellent type~I error control
when the underlying network topology is preserved, with mean $p$-values near~1 across all
sample sizes. In contrast, under differential lesion conditions, the topological transposition
test and T--ANOVA consistently achieve near-zero $p$-values, demonstrating high sensitivity
to lesion-induced topological changes. Heo's ANOVA frequently fails to detect these effects
and exhibits substantially larger variability, particularly in smaller samples. PERMANOVA
performs better than Heo's ANOVA but remains less sensitive than the proposed topological
methods. Overall, the error patterns indicate that the topological procedures are both
statistically stable under noise and more powerful in identifying meaningful differences in
network topology. Furthermore, computational time compared across the three ANOVA methods are shown in Figure~\ref{fig: sim_time}. As expected, the computational time of T-ANOVA grew steadily in comparison to the exponential growth of the other two methods.

\section{Application}
\label{sec: application}

Stroke is the leading cause of severe adult disability in the United States \citep{Tsao2022}. A left-hemisphere stroke commonly leads to aphasia, a speech-language disorder often classified into subtypes according to behavioral symptoms using tests such as the Western Aphasia Battery (WAB). 

\subsection{Data acquisition and preprocessing} 
\label{sec: data}

The resting-state fMRI (rs-fMRI) data were acquired from 96 participants with anomic or Broca's aphasia resulting from a single ischemic or hemorrhagic stroke involving the left hemisphere on a Siemens Prisma 3T scanner with a 20-channel head coil located at the Center for the Study of Aphasia Recovery (C-STAR) at the University of South Carolina \citep{Riccardi2024}. The neuroimaging scans were approved by the University Institutional Review Board (IRB). The following imaging parameters were used: a multiband sequence (x2) with a $216 \times216$ mm field of view, a $90 \times 90$ matrix size, and a 72-degree flip angle, 50 axial slices (2 mm thick with $20 \% $ gap yielding 2.4 mm between slice centers), repetition time TR = 1650 ms, TE = 35 ms, GRAPPA = 2, 44 reference lines, interleaved ascending slice order. During the scanning process, the participants were instructed to stay still with eyes closed. A total of 370 volumes were acquired. The preprocessing procedures of the rs-fMRI data include motion correction, brain extraction and time correction using a novel method developed for stroke patients \citep{Yourganov2018}. The Realign and Unwarp procedure in SPM12 with default settings was used for motion correction. Brain extraction was then performed using the SPM12 script pm\_brain\_mask with default settings. Slice time correction was also done using SPM12. The mean fMRI volume for each participant was then aligned to the corresponding T2-weighted image to compute the spatial transformation between the data and the lesion mask. The fMRI data were then spatially smoothed with a Gaussian kernel with FWHM= 6 mm. To eliminate artifacts driven by lesions, a pipeline proposed by \citet{Yourganov2018} was applied on the rs-fMRI. The FSL MELODIC package was used to decompose the data into independent components (ICs) and to compute the Z-scored spatial maps for the ICs. The spatial maps were thresholded at $p < 0.05$ and compared with the lesion mask for the participant. The Jaccard index, computed as the ratio between the numbers of voxels in the intersection and union, was used to quantify the amount of spatial overlap between the lesion mask and thresholded IC maps, both of which were binary. ICs corresponding to Jaccard index greater than $5\%$ were deemed significantly overlapping with the lesion mask and then regressed out of the fMRI data using the fsl\_regfilt script from the FSL package. By applying the automated anatomical labelling (AAL) and Johns Hopkins University (JHU) atlases, respectively 116 (in application) and 189 (in simulations) regions of interest (ROIs) were created and used as nodes in the resting-state functional brain networks subsequently constructed. Pearson's correlation corrected for stroke lesion presence were used as the edge weights of the networks. In terms of behavioral measures, the following revised WAB (WAB-R) subscores \citep{Kertesz2007} were used to measure performance of participants in fluency, repetition, comprehension, and naming. We note that only the first three categories play a role in determining the traditional aphasic subtypes. But we included naming as it represents an important aspect of assessing aphasic impairment. 

\begin{figure}[b!]
\centering
  \includegraphics[width = 0.8\linewidth, trim = 0in 0.5in 0in 0in]{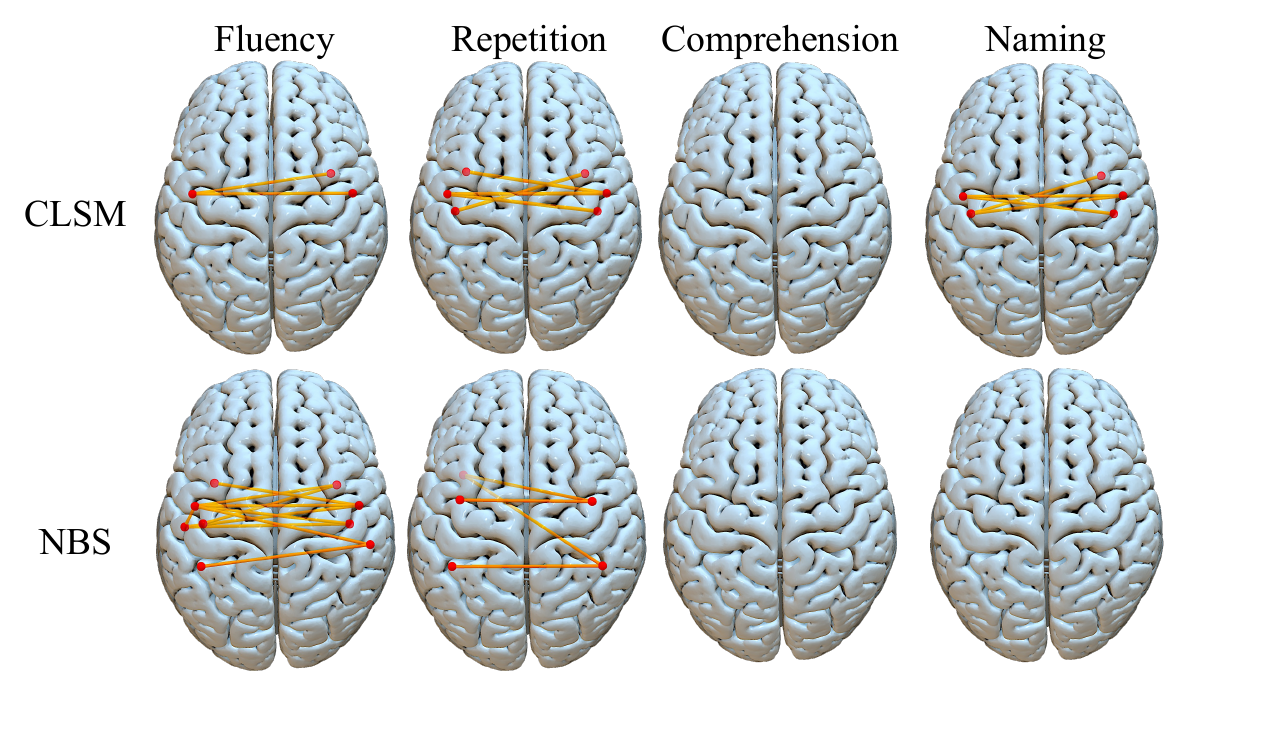}
  \caption{\label{fig: lsm}Results of CLSM and NBS: connections that distinguish between the lower and higher subscore groups.}
\end{figure}

\subsection{Standard lesion symptom mapping} Standard CLSM methods typically identify a number of edges, or none, that distinguish two groups of brain networks. To establish a baseline, we modified standard LSM analyses carried out in \citep{Riccardi2024}, where two traditional mass univariate methods: voxel- , region- and connectivity-based LSM (VLSM, RLSM, CLSM), were applied to compare post-stroke anomic and Broca's aphasia groups in the C-STAR dataset. Findings include damage to the ventral precentral gyrus was associated with Broca’s aphasia (non-fluent) compared to a group with anomic aphasia, who generally have preserved fluency. Here, we use lower and higher 50\% subscores as the new grouping in the LSM analyses and a related network-based statistic (NBS) inference method identifying a subnetwork distinguishing two groups of brain networks, first applied in an LSM setting by \citet{Riccardi2024}. Results of the CLSM and NBS comparing between the lower and higher subscore groups of fluency, repetition, comprehension, and naming are shown in Figure~\ref{fig: lsm}. We can see that CLSM and NBS tend to yield none or very few connections. For instance, there was no significant connection identified for the comprehension group comparison and only one connection between Rolandic\_Oper\_L and Rolandic\_Oper\_R and one between Rolandic\_Oper\_L and Insula\_R for the naming group comparison by CLSM. There were no significant connection identified for both of these categories by NBS. We also applied to the C-STAR data the more recent machine learning-based multivariate LSM (MLSM), but the results were either nonconvergent or subpar with accuracies close to 50\% —  most likely due to the low sample sizes compared with the hundreds of participants suggested by recent studies \citep{Karnath2018, Ivanova2021}. This in turn shows another important advantage of our TLSM approach in terms of not requiring large sample sizes typically only available in consortium-style datasets.

\subsection{Topological lesion symptom mapping} As we mentioned earlier, TLSM is not aimed at "outperforming" any of the existing LSM approaches. It is meant to fills a gap in connectome/network-based inference approach beyond the order of single connections. Instead of thresholding an edge-level statistic to identify connections of interest like the standard LSM and NBS methods, our TLSM approach tracks cycles formed of multiple connections through all possible thresholds on the edge weights and encodes the thresholds at which the cycles appear and die in a PD for subsequent inference. In detail, each PD used in TLSM is derived from the Rips filtration of an individual's resting-state functional connectivity matrix, so the topological features under comparison reflect
mesoscale properties of the underlying brain networks. We highlight this connection to keep the two-sample transposition test and T-ANOVA  anchored in the LSM framework: any detected group difference corresponds to a difference in the persistence of network cycles across behavioral subgroups. After constructing the PDs and smoothing them with the HK representation, we apply either the two-sample transposition test or T-ANOVA, depending on the number of groups, and interpret significant findings by examining the connectivity patterns of the underlying cycles.

An important consideration we had in setting up this TLSM framework, first of its kind, is consistency in topological structures across brain networks. It is expected that the brain network topology alters significantly across individuals with heterogeneous lesion presence and thus certain cycles may not be consistently observed across brain networks of all, or even most, individuals. In this study, we 
filtered out resting-state functional brain networks with 1-cycles of interest and carry out the topological inference procedure on their persistence patterns. Note that the approach of focusing on 1-cycles in the form polygons of different orders was also adapted by recent works in neuroscience to study topological signatures in the structural brain networks of neurotypical subjects \citep{Sizemore2018}. In our application to resting-state functional brain networks from individuals with post-stroke aphasia, we further narrowed down the polygons of order $K$ ($K \ge 4$) through three combinatorial conditions imposed on 36 ROI nodes from three lobes (frontal/parietal/temporal) known to be involved in language processing: 
\begin{itemize}
  \item LK: all $K$ nodes in the polygon are from the same lobe;
  \item LK1: $K-1$ nodes in the polygon are from the same lobe;
  \item LK2: $K-2$ nodes in the polygon are from the same lobe. 
 \end{itemize}
There are no other possibilities because of the three-lobe constraints. Such 1-cycles we identified from the C-STAR dataset per nodes created from the AAL atlas were mostly polygons with four sides or 4-polygons: 17 and 34 respectively satisfying the combinatorial conditions LK1 and LK2, and none for LK. So we simply focused on comparing the persistence patterns of these 4-polygons in two or more groups through our proposed topological inference procedure. The grouping we used in this study was participants with lower vs upper 50\% WAB subscores in the four categories fluency, repetition, comprehension, and naming for two-group comparison, as well as lower vs middle vs upper third subscores for multi-group comparison. In general, the framework can be applied to persistence patterns of any assortment of 1-cycles decoded from the brain networks of groups under comparison. 

\begin{figure}[t!]
  \includegraphics[width = 1\linewidth]{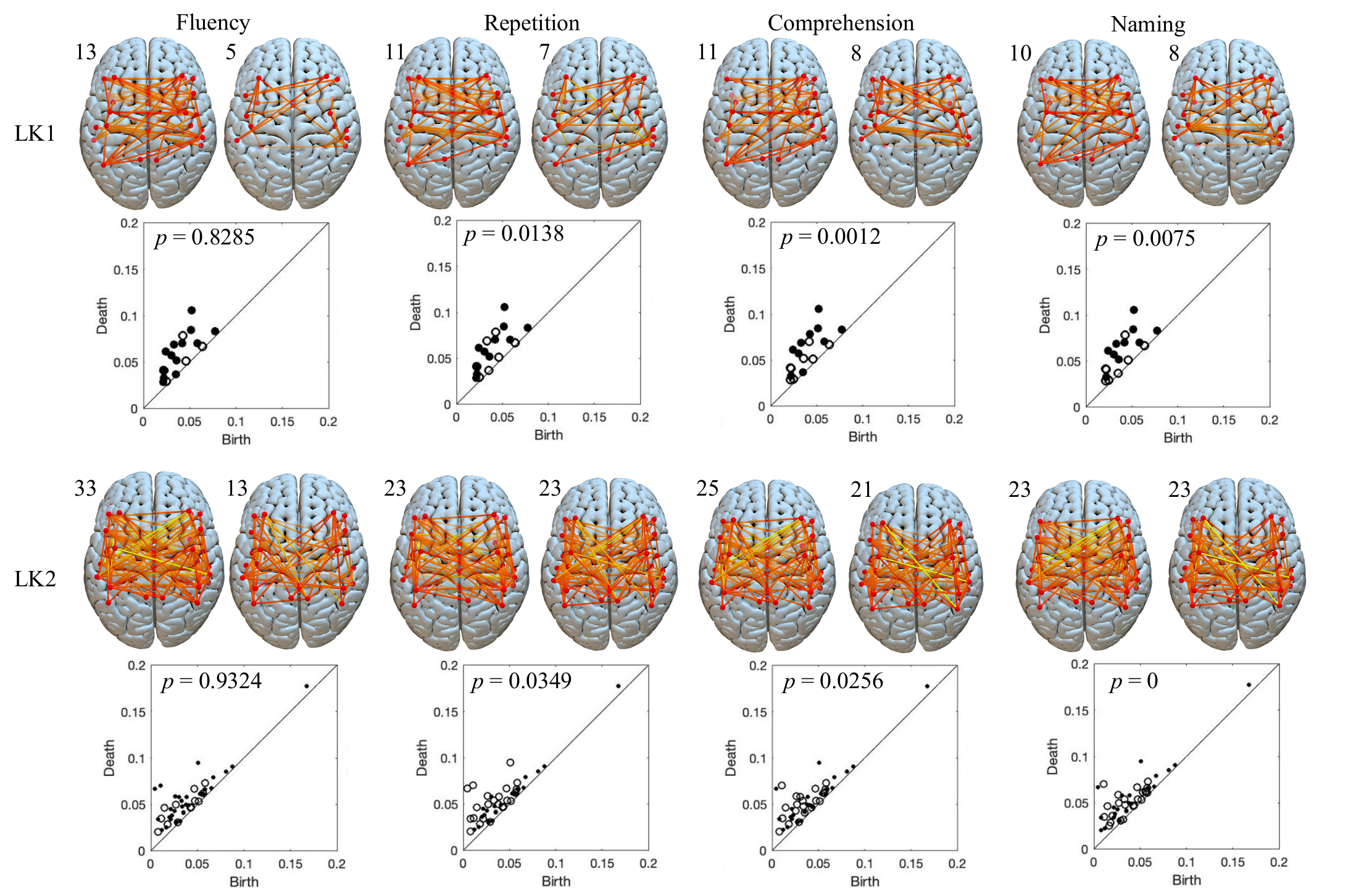}
  \caption{\label{fig: tlsm} The LK1 and LK2 4-polygons collected from resting-state functional brain networks of subjects with the lower (left of each pair) and upper (right of each pair) 50\% WAB subscores in the four categories fluency, repetition, comprehension, and naming, as well as the $p$-values from the two-sample transposition test comparing the persistence patterns of the 4-polygons (dot: lower 50\% subscores, circle: higher 50\% subscores). The numbers on the upper left corners of the connectivity maps pertain to the 4-polygons in the respective groups.}
\end{figure}

Figure~\ref{fig: tlsm} shows the LK1 and LK2 4-polygons collected from resting-state functional brain networks of subjects with the lower and upper 50\% WAB subscores in the four categories fluency, repetition, comprehension, and naming, as well as the $p$-values comparing the persistence patterns of the 4-polygons. There was a heavy overlap between subjects with the lower 50\% subscores across the four categories. The LK1 4-polygons make up distinctly different connectivity patterns between the lower and higher subscore groups across all four categories, with the lower subscore groups appearing much "denser" than the higher subscore groups because of a larger number of these polygons across all four categories. So the connectivity patterns are similar in all of the lower or higher score groups although the latter do show more visual difference than the former, most likely down to the sparseness in comparison. Meanwhile, the LK2 4-polygons are more comparable in group sizes and also connect similar nodes between the lower and higher subscore groups across all four categories. It is thus interesting to see a polygon with just one node altered would end up with a lower score group. For instance, in the naming category, subject M2084 in the higher subscore group has a 4-polygon connecting SupraMarginal\_R, Temporal\_Sup\_R, Temporal\_Inf\_R, and Precentral\_L, whereas subject M2016 in the lower subscore group has a 4-polygon connecting SupraMarginal\_R, Temporal\_Sup\_R, Temporal\_Inf\_R, and Precentral\_R. It is worth noting in this case that the node that seems to make the difference is in the left hemisphere of the subject with a higher score, whereas all four nodes in the polygon of the subject with lower subscore are all from the right hemisphere. So even though the polygons only differ by a node, it could be that the former is a "favorable" structure. 

On the other hand, the persistence patterns provide one more layer of information to the cycles. Inference is also conducted at the level of the persistence patterns. We can see that the persistence patterns in the fluency category do not differ significantly between the lower and higher subscore groups ($p$-values from the two-sample transposition test are 0.8285 and 0.9324 for the LK1 and LK2 4-polygons respectively). Comprehension and naming stand out as the most significant out of the LK1 and LK2 comparisons respectively ($p$-values = 0.0012, 0). On closer inspection we found that inference for several 4-polygons is most sensitive to the reassignment of subjects who exhibit relatively high persistence values for those polygons. In other words, moving a subject that carries a highly persistent loop from one behavioral group to the other has a larger impact on the permutation distribution — and hence the p-value — than moving subjects with low-persistence values. Biologically, high-persistence points mark robust, redundant loops that are likely to represent resilient integrative circuits; when these highly persistent polygons segregate preferentially into one behavioral group, the result plausibly reflects a real difference in the availability of robust mesoscale pathways between groups. We therefore interpret significant TLSM findings driven by such high-persistence reassignment as potentially meaningful indicators of preserved (or lost) integrative circuitry rather than as artifacts of marginal noise. To complement the two-group comparison, we applied T-ANOVA to compare the persistence patterns between three groups of subscores (lower, middle, and upper third) across the four categories. But we did not detect any significant difference, most likely due to the tiny group sizes when split three ways.

\section{Discussion}

In this study, we developed a topological inference framework based on the heat-kernel (HK) representation of persistence diagrams (PDs). Because inference is carried out directly on HK coefficients, the framework is agnostic to the type of data from which the PDs are derived. In this paper, we centered the application on group comparisons of PDs extracted from brain networks, where homological features naturally reflect mesoscale connectivity structure. Methodologically, the framework addresses several gaps left by prior work. The proposed spectral transposition test extends the permutation scheme of \citet{Wang2018}, which was limited to single-trial univariate signals and fixed Fourier bases. By performing transpositions directly on HK representations of PDs, we enable fast, incremental permutation updates for multi-trial, multivariate, and network-based persistence features, without requiring alignment or matching of points across diagrams. This is particularly important because PDs are inherently heterogeneous and lack unique statistical means; the HK representation resolves this challenge and makes rigorous resampling-based inference feasible for high-dimensional topological features. We further introduced a topological ANOVA (T-ANOVA) framework for multi-group comparisons of persistence features without dimensionality reduction. Unlike projection-based approaches such as \citet{Heo2012}, which may distort homological structure, T-ANOVA operates directly on precomputed HK distances and updates them efficiently under transposition. The result is a fully nonparametric, topology-preserving ANOVA procedure for complex, non-Euclidean data. Beyond its topological setting, the multi-group transposition mechanism provides a general strategy for accelerating permutation-based inference. In addition, HK coefficients offer a principled basis for resampling persistence features, suggesting potential extensions to data augmentation strategies in deep learning, particularly in small-sample regimes \citep{Huang2021}.

Beyond methodological contributions, our simulation design is also novel. Simulating brain networks with cycles is not straightforward, so we used point clouds simulated from images with an underlying keyshape and networks mimicking real functional brain connectivity to study the robustness and sensitivity of the proposed methods. Since we did not develop an automated bandwidth selection method in this study, where we applied the bandwidth $\sigma = 10$ throughout simulations and application, we conducted stability analysis of the T-ANOVA method under different bandwidths in several simulation settings. The results, included in the Appendix, remained stable when we varied the bandwidth to $\sigma = 5$.

Yet another key contribution of our work is demonstrating how the proposed inference framework enables a new form of lesion symptom mapping (LSM) based on topological features of brain networks. Traditional LSM and connectome-based LSM focus on identifying voxels, regions, or individual connections whose damage is associated with behavioral impairment. Our topological LSM (TLSM) procedure, by contrast, evaluates mesoscale connectivity motifs—specifically, the persistence of 1-cycles that form robust loops linking frontal, temporal, and parietal regions central to language processing. By applying TLSM to resting-state functional brain networks from individuals with post-stroke aphasia, we were able to characterize how the robustness of these polygonal structures varies with fluency, repetition, comprehension, and naming performance. The results suggest that differences in aphasic impairment are reflected not only in focal disconnections but also in how resilient or fragile certain integrative network loops become following a stroke. This illustrates the value of TLSM as a complementary tool for clinical neuroscience, offering insight into the topological organization of brain networks beyond edge- or region-based analyses.

Finally, although this paper focuses on topological inference, the same principles extend naturally to a clustering framework. In an earlier version of this paper \citep{Wang2023}, we demonstrated distance-based topological clustering for subtyping post-stroke aphasia, illustrating the utility of T-ANOVA for identifying clinically meaningful variability in network topology. These developments, along with additional methodological refinements and longitudinal extensions, are pursued in companion studies \citep{Yi2025, Yi2026IEEE}. 

Taken together, these contributions establish a coherent and flexible statistical framework for inference on persistence diagrams that did not previously exist. The novelty lies in adapting HK representations for hypothesis testing, unifying fast permutation and
multi-group inference into a single topological pipeline, and introducing TLSM, which extends the traditional LSM paradigm by enabling group comparison of persistence features to detect mesoscale network disruptions. This integration of topological inference with a clinically established framework highlights the methodological and translational significance of the proposed approach.

\section*{Acknowledgments} We thank Dr. Moo K. Chung for helpful discussions on the methods. Funding sources: NIHP50DC014664 (PI: JF, Project PI: DDO), NIH R01DC017162 and R01DC01716202S1 (PI: RHD). Author contributions: conceptualization (YW, JY, NR,  RHD), statistical analysis (YW, JY, NR), interpretation of results (All), writing and editing (All). 

\bibliographystyle{plainnat}
\bibliography{bib_top_inf_aoas} 

\clearpage
\setcounter{page}{1}
\renewcommand{\thepage}{S\arabic{page}}
\newpage

\clearpage

\renewcommand{\thefigure}{S\arabic{figure}}
\renewcommand{\thetable}{S\arabic{table}}
\captionsetup[figure]{name=Figure}
\captionsetup[table]{name=Table}
\setcounter{figure}{0}
\setcounter{table}{0}

\section*{Appendix} Additional details to the main text are included here.\\

\noindent{\em Heat kernel smoothing on a general manifold.} We assume the fundamental stochastic model
\begin{equation}
\label{eq: model}
f(p) = h(p) + \varepsilon(p), p\in\mathcal{M},
\end{equation}
where $f$ is the noisy measurement, $h$ is the unknown signal, and $\varepsilon$ is a zero-mean Gaussian random field. We make the general enough assumptions that $f\in L_2(\mathcal{M})$, the space of square integrable  functions on $\mathcal{M}$ with the inner product $\langle f_1,f_2\rangle=\int_{\mathcal{M}}f_1(p)f_2(p)d\mu(p)$, where $\mu$ is the Lebesgue measure. A self-adjoint operator $\mathcal{L}$, i.e. $\langle f_1,\mathcal{L}f_2\rangle = \langle \mathcal{L}f_1,f_2\rangle$ for all $f_1, f_2\in L_2(\mathcal{M})$, induces orthonormal eigenvalues $\lambda_k$ and eigenfunctions $\psi_k$ on $\mathcal{M}$: 
\begin{equation}
\label{eq: eigen}
\mathcal{L}\psi_k = \lambda_k\psi_k, k = 0, 1, \dots,
\end{equation}
where, without loss of generality, we can sort the eigenvalues $\lambda_k$ such that $$0 = \lambda_0 \le \lambda_1 \le \cdots,$$ and the eigenfunctions $\psi_k$ can be numerically computed by solving a generalized eigenvalue problem. Then, by Mercer's Theorem, any symmetric positive definite kernel can be written as
\begin{equation}
\label{eq: spd}
K(p,q) = \sum_{k=0}^{\infty}\tau_k\psi_k(p)\psi_k(q)
\end{equation}

Now consider the diffusion-like Cauchy problem
\begin{equation}
\label{eq: diffusion}
\frac{\partial h(\sigma, p)}{\partial \sigma} + \mathcal{L}(h(\sigma, p)) = 0, p\in\mathcal{M},
\end{equation}
with the initial condition $h(\sigma = 0, p) = f(p)$. The partial differential equation \eqref{eq: diffusion} diffuses the noisy data $h(p)$ over $\sigma$. For the self-adjoint operator $\mathcal{L}$, \eqref{eq: diffusion} has the unique solution \citepsupp{Chung2007}
\begin{equation}
\label{eq: diffusion_sol}
h(\sigma,p)=\sum_{k = 0}^{\infty}e^{-\lambda_k\sigma}\langle h,\psi_k\rangle\psi_k(p),
\end{equation}
which provides an estimate $\hat{h}_{\sigma}(p)$ of the unknown signal $h(p)$. The bandwidth $\sigma$ controls the amount of smoothing in the estimate; as $\sigma$ increases, $\hat{h}_{\sigma}(p)$ becomes smoother. When $\mathcal{L}$ is the Laplace-Beltrami (LB) operator, the diffusion equation \eqref{eq: diffusion} becomes the isotropic heat diffusion equation and the kernel \eqref{eq: spd} becomes the heat kernel (HK)
\begin{equation}
\label{eq: hk}
K_{\sigma}(p,q) = \sum_{k=0}^{\infty}e^{-\lambda_k\sigma}\psi_k(p)\psi_k(q), p, q\in\mathcal{M},
\end{equation}
where the $\psi_k$ are the eigenfunctions of the LB operator $\Delta$ satisfying
$\Delta \psi_k(p) = \lambda_k \psi_k(p)$
for $p \in \mathcal{M}$. The HK framework has been shown to be equivalent to kernel regression and wavelet \citepsupp{Chung2014.MICCAI}. \\

\noindent{\em Simulation studies based on point clouds.} 
Tables~\ref{tab: results_trans_robustness}, \ref{tab: results_anova_power} and \label{tab: results_anova_robustness} summarize the results of the point cloud simulations
 \begin{table}[h!]
  \scalebox{0.6}{
  \begin{tabular}{c|cc|cc}
  \hline
  & \multicolumn{4}{c}{{\bf Robustness under Variation of Topological Noise}}\\\hline
  & \multicolumn{2}{c|}{Number of points: 100} & \multicolumn{2}{c}{Number of points: 200}\\\hline
  Percentage  & Pre-specified Loc. ($m = 4$ vs $n = 4$) & Random Loc. ($m = 5$ vs $n = 5$) & Pre-specified Loc. ($m = 4$ vs $n = 4$) & Random Loc. ($m = 5$ vs $n = 5$)\\\hline
  100\% & 0.4567$\pm$0.2874 & 0.4133$\pm$0.2482 & $0.5071 \pm 0.2923$ & $0.5322 \pm 0.2792$\\
  95\% & 0.4777$\pm$0.2843 & 0.4498$\pm$0.2844 & $0.4795 \pm 0.3038$ & $0.5183 \pm 0.2757$\\
  90\% & 0.4455$\pm$0.2791 & 0.5214$\pm$0.2983 & $0.5203 \pm 0.2725$ & $0.4687 \pm 0.2764$\\\hline
    Percentage  & Random Loc. ($m = 20$ vs $n = 20$) & Random Loc. ($m = 100$ vs $n = 100$) & Random Loc. ($m = 20$ vs $n = 20$) & Random Loc. ($m = 100$ vs $n = 100$)\\\hline
  100\% & 0.5060$\pm$0.3163 & 0.4328$\pm$0.2764 & $0.4942 \pm 0.3090$ & $0.5465 \pm 0.2934$\\
  95\% & 0.5016$\pm$0.2998 & 0.4193$\pm$0.2863 & $0.4817 \pm 0.2951$ &  $0.5344 \pm 0.3014$\\
  90\% & 0.4827$\pm$0.2919 & 0.5260$\pm$0.2812 & $0.5336 \pm 0.3123$ & $0.5441 \pm 0.3052$\\\hline
  & \multicolumn{4}{c}{{\bf Robustness under Variation of Cycle Location}} \\\hline
  & \multicolumn{2}{c|}{Number of points: 100} & \multicolumn{2}{c}{Number of points: 200}\\\hline
   Percentage & Pre-specified Loc. ($m = 4$ vs $n = 4$) & Random Loc. ($m = 5$ vs $n = 5$) & Pre-specified Loc. ($m = 4$ vs $n = 4$) & Random Loc. ($m = 5$ vs $n = 5$)\\\hline
  100\% & 0.2917$\pm$0.2624 & 0.5005$\pm$0.2883 & $0.4516 \pm 0.2876$ & $0.4566 \pm 0.2907$\\
  95\% & 0.2973$\pm$0.2407 & 0.5342$\pm$0.2775 & $0.4469 \pm 0.2960$ & $0.4750 \pm 0.2900$\\
  90\% & 0.3065$\pm$0.2505 & 0.4434$\pm$0.3050 & $0.5658 \pm 0.2995$ & $0.5327 \pm 0.2877$\\\hline
  Percentage & Radom Loc. ($m = 20$ vs $n = 20$) & Random Loc. ($m = 100$ vs $n = 100$) & Radom Loc. ($m = 20$ vs $n = 20$) & Random Loc. ($m = 100$ vs $n = 100$)\\\hline
  100\% & 0.4998$\pm$0.2901 & 0.3845$\pm$0.2620 & $0.5173 \pm 0.2823$ & $0.4970 \pm 0.2769$\\
  95\% & 0.4608$\pm$0.2999 & 0.3924$\pm$0.2777 & $0.4797 \pm 0.2813$ & $0.5165 \pm 0.3088$\\
  90\% & 0.4810$\pm$0.2568 & 0.4550$\pm$0.2835 & $0.5085 \pm 0.2828$ & $0.5463 \pm 0.2839$\\\hline
  \end{tabular}}
  \caption{\label{tab: results_trans_robustness}Summary of mean$\pm$standard deviation of $p$-values from the spectral transposition test in 100 simulations. Top: In each simulation, the test was used to compare a group of $m$ random samples with a varied percentage (90\%, 95\%, 100\%) of 100 or 200 points from the original key shape with a group of $n$ random samples with the same percentage of 100 or 200 points from the key shape with topological noise in the form of a much smaller 1-cycle next to the keyhole. The location of the smaller cycle in each random sample of the second group was also replaced with pre-specified options or randomly chosen from the pre-specified options. Bottom: In each simulation, the test was used to compare a group of $m$ random samples with a varied percentage (90\%, 95\%, 100\%) of 100 or 200 points from the original key shape with only the top left quarter of the keyhole left, with a group of $n$ random samples with the same percentage of 100 or 200 points from the original key shape with a random quarter of the keyhole left.}
  \end{table}

\begin{table}[h!]
\scalebox{0.6}{
  \begin{tabular}{c|ccc|ccc}
  \hline
   \multicolumn{7}{c}{{\bf Sensitivity in Detecting Differential cycle Presence Among Multiple Groups}} \\\hline
   & \multicolumn{3}{c|}{Number of points: 100} & \multicolumn{3}{c}{Number of points: 200}\\\hline
   & \multicolumn{6}{c}{$n_1 = n_2 = n_3 = 5$} \\\hline
    Percentage of Points in Key Shape & T-ANOVA & Heo's ANOVA & PERMANOVA & T-ANOVA & Heo's ANOVA & PERMANOVA \\\hline
  100\% & $0.0009 \pm 0.0016$ & $0.0038 \pm 0.0101$    & $0.0025 \pm 0.0033$ & $0.0004\pm0.0003$ & $0.0000\pm0.0000$ & $0.0004\pm0.0004$\\
  95\% & $0.0016 \pm 0.0033$ & $0.0109 \pm 0.0258$     & $0.0045 \pm 0.0064$ & $0.0005\pm0.0003$ & $0.0000\pm0.0000$ & $0.0005\pm0.0006$\\
  90\% & $0.0060 \pm 0.0146$ & $0.0436 \pm 0.0791$     & $0.0192 \pm 0.0438$ & $0.0005\pm0.0003$ & $0.0001\pm 0.0003$ & $0.0006\pm0.0009$\\ \hline
    & \multicolumn{6}{c}{$n_1 = n_2 = n_3 = 20$}\\\hline
    Percentage of Points in Key Shape & T-ANOVA & Heo's ANOVA & PERMANOVA & T-ANOVA & Heo's ANOVA & PERMANOVA \\\hline
  100\% & $0.0000 \pm 0.0000$ & $0.0000 \pm 0.0000$	& $0.0000 \pm 0.0000$ & $0.0000\pm0.0000$ & $0.0000\pm0.0000$ & $0.0000\pm0.0000$\\
  95\% & $0.0000 \pm 0.0000$ & $0.0000 \pm 0.0000$	& $0.0000 \pm 0.0000$ & $0.0000\pm0.0000$ & $0.0000\pm0.0000$ & $0.0000\pm0.0000$\\
  90\% & $0.0000 \pm 0.0000$ & $0.0000 \pm 0.0001$	& $0.0000 \pm 0.0000$ & $0.0000\pm0.0000$ & $0.0000\pm0.0000$ & $0.0000\pm0.0000$\\ \hline
    & \multicolumn{6}{c}{$n_1 = 5, n_2 = 20, n_3 = 100$} \\\hline
    Percentage of Points in Key Shape & T-ANOVA & Heo's ANOVA & PERMANOVA & T-ANOVA & Heo's ANOVA & PERMANOVA \\\hline
  100\% & $0.0253 \pm 0.0699$ & $0.0353 \pm 0.0101$   & $0.0011 \pm 0.0055$ & $0.0000\pm0.0001$ & $0.0000\pm 0.0000$ & $0.0000\pm0.0000$\\
  95\% & $0.0485 \pm 0.0827$ 	& $0.0511 \pm 0.1255$   & $0.0028 \pm 0.0117$ & $0.0000\pm0.0001$ & $0.0000\pm 0.0000$ & $0.0000\pm0.0000$\\
  90\% & $0.0998 \pm 0.1377$ 	& $0.1281 \pm 0.2395$   & $0.0091 \pm 0.0238$ & $0.0001\pm0.0004$ & $0.0000\pm 0.0000$ & $0.0000\pm0.0000$\\ \hline
  \end{tabular}}
  \caption{\label{tab: results_anova_power} Summary of mean$\pm$standard deviation of $p$-values 
  from the T-ANOVA, Heo's ANOVA, and PERMANOVA in 100 simulations. In each simulation, three groups of $n_1, n_2, n_3$ 100- or 200-point point clouds were generated, where the 100 or 200 points in each point cloud of the first two groups are generated randomly from the part of the rectangular image, whereas the 100 or 200 points in each point cloud of the third group are generated randomly with a varied percentage (90\%, 95\%, 100\%) of points from the shape of the key.}
  \end{table}

\begin{table}[h!]
\scalebox{0.6}{
  \begin{tabular}{c|ccc|ccc}
  \hline
   & \multicolumn{6}{c}{{\bf Robustness under Variation of Topological Noise}} \\\hline
   & \multicolumn{3}{c|}{Number of points: 100} & \multicolumn{3}{c}{Number of points: 200}\\\hline
   & \multicolumn{6}{c}{Pre-specified Location: $n_1 = n_2 = n_3 = 4$}\\\hline
   Percentage of Points in Key Shape & T-ANOVA & Heo's ANOVA & PERMANOVA  & T-ANOVA & Heo's ANOVA & PERMANOVA\\\hline
  100\% & $0.4860 \pm 0.2607$ & $0.5267 \pm 0.2675$   & $0.4901 \pm 0.2965$ & $0.5058\pm0.3116$ & $0.5390\pm0.2850$ & $0.5087\pm0.2898$\\
  95\% & $0.4897 \pm 0.2608$ 	& $0.5211 \pm 0.2822$   & $0.5050 \pm 0.2851$ & $0.4845\pm0.3094$ & $0.4545\pm0.3115$ & $0.5044\pm0.3012$ \\
  90\% & $0.4974 \pm 0.2963$  & $0.5163 \pm 0.3125$   & $0.4619 \pm 0.2667$ & $0.5265\pm0.2690$ & $0.4904\pm0.2948$ & $0.5007\pm0.2834$\\ \hline
  & \multicolumn{6}{c}{Random Location: $n_1 = n_2 = n_3 = 5$} \\\hline
     Percentage of Points in Key Shape & T-ANOVA & Heo's ANOVA & PERMANOVA  & T-ANOVA & Heo's ANOVA & PERMANOVA\\\hline
  100\% & $ 0.5275 \pm 0.3052$ & $0.4988 \pm 0.2882$   & $0.5518 \pm 0.2952$ & $0.5634\pm0.2713$ & $0.5182\pm0.2913$ & $0.5236\pm0.2938$ \\
  95\% & $0.5166 \pm 0.3138$  & $0.5062 \pm 0.2830$    & $0.5308 \pm 0.2870$ & $0.5535\pm0.2682$  & $0.4798\pm0.2860$ & $0.5205\pm0.2941$\\
  90\% & $0.4970 \pm 0.2975$	& $0.5205 \pm 0.2689$    & $0.4923 \pm 0.2928$ & $0.4836\pm0.2732$ & $0.5212\pm0.3035$ & $0.5067\pm0.2762$\\ \hline
  & \multicolumn{6}{c}{Random Location: $n_1 = n_2 = n_3 = 20$}\\\hline
     Percentage of Points in Key Shape & T-ANOVA & Heo's ANOVA & PERMANOVA & T-ANOVA & Heo's ANOVA & PERMANOVA \\\hline
  100\% & $0.4915 \pm 0.2624$	& $0.4812 \pm 0.2640$   & $0.4727 \pm 0.2720$ & $0.4882\pm0.2975$ & $0.5152\pm0.3271$ & $0.4691\pm0.2727$\\
  95\% & $0.4860 \pm 0.2862$  & $0.4864 \pm 0.3072$   & $0.5141 \pm 0.2734$ & $0.4649\pm0.3022$ & $0.4246\pm0.2849$ & $0.4493\pm0.2800$\\
  90\% & $0.4349 \pm 0.2759$  & $0.5139 \pm 0.2791$   & $0.4706 \pm 0.2996$ & $0.5440\pm0.3060$ & $0.5214\pm0.3129$ & $0.5318\pm0.2940$\\ \hline
  & \multicolumn{6}{c}{Random Location: $n_1 = 5, n_2 = 20, n_3 = 100$} \\\hline
     Percentage of Points in Key Shape & T-ANOVA & Heo's ANOVA & PERMANOVA & T-ANOVA & Heo's ANOVA & PERMANOVA \\\hline
  100\% & $0.5374 \pm 0.2937$ & $0.5001 \pm 0.3028$   & $0.4593 \pm 0.2836$ & $0.4547\pm0.2932$ & $0.4599\pm0.2825$ & $0.4840\pm0.3076$\\
  95\% & $0.5413 \pm 0.2747$  & $0.5016 \pm 0.2738$   & $0.4692 \pm 0.3037$ & $0.4254\pm0.2813$ & $0.4783\pm0.2879$ & $0.5134\pm0.2835$\\
  90\% & $0.4938 \pm 0.2855$ 	& $0.5014 \pm 0.2955$   & $0.5085 \pm 0.3040$ & $0.4500\pm0.3001$ & $0.4731\pm0.3114$ & $0.4455\pm0.2898$\\ \hline
 & \multicolumn{6}{c}{{\bf Robustness under Variation of cycle Location}} \\\hline
    & \multicolumn{3}{c|}{Number of points: 100} & \multicolumn{3}{c}{Number of points: 200}\\\hline
  & \multicolumn{6}{c}{Pre-specified Location: $n_1 = n_2 = n_3 = 4$}\\\hline
   Percentage of Points in Key Shape & T-ANOVA & Heo's ANOVA & PERMANOVA & T-ANOVA & Heo's ANOVA & PERMANOVA\\\hline
  100\% & $0.4887 \pm 0.2873$   & $0.4664 \pm 0.2537$   & $0.5057 \pm 0.2862$ & $0.5092\pm0.2915$ & $0.4988\pm0.2953$ & $0.4838\pm0.2840$\\
  95\% & $0.4948 \pm 0.2776$ 	  & $0.4479 \pm 0.2889$   & $0.5109 \pm 0.2779$ & $0.4857\pm0.3033$ & $0.4964\pm0.2705$ & $0.4834\pm0.2897$\\
  90\% & $0.4364 \pm 0.2572$	  & $0.4651 \pm 0.2916$   & $0.4463 \pm 0.2767$ & $0.5255\pm0.2854$ & $0.5565\pm0.2873$ & $0.4590\pm0.2835$\\ \hline
  & \multicolumn{6}{c}{Random Location: $n_1 = n_2 = n_3 = 5$} \\\hline
   Percentage of Points in Key Shape & T-ANOVA & Heo's ANOVA & PERMANOVA & T-ANOVA & Heo's ANOVA & PERMANOVA\\\hline
  100\% & $0.4785 \pm 0.2737$	 & $0.5505 \pm 0.2706$   & $0.4932 \pm 0.3036$ & $0.4710\pm0.2901$ & $0.4672\pm0.2829$ & $0.4740\pm0.2888$\\
  95\% & $0.5039 \pm 0.3082$ 	 & $0.4872 \pm 0.3055$   & $0.4572 \pm 0.3079$ & $0.5146\pm0.2906$ & $0.5337\pm0.2730$ & $0.4997\pm0.2871$\\
  90\% & $0.4887 \pm 0.2919$   & $0.5302 \pm 0.2784$   & $0.4095 \pm 0.2784$ & $0.5173\pm0.3079$ & $0.4716\pm0.2956$ & $0.4563\pm0.2846$ \\ \hline
  & \multicolumn{6}{c}{Random Location: $n_1 = n_2 = n_3 = 20$}\\\hline
   Percentage of Points in Key Shape & T-ANOVA & Heo's ANOVA & PERMANOVA & T-ANOVA & Heo's ANOVA & PERMANOVA\\\hline
  100\% & $0.5249 \pm 0.3118$ & $0.4839 \pm 0.2928$   & $0.5048 \pm 0.3159$ & $0.5091\pm0.2952$ & $0.5279\pm0.2660$ & $0.4938\pm0.2766$\\
  95\% & $0.5292 \pm 0.3082$  & $0.4528 \pm 0.2790$   & $0.5398 \pm 0.3044$ & $0.4869\pm0.2726$ & $0.5377\pm0.2870$ & $0.5201\pm0.2860$ \\
  90\% & $0.5183 \pm 0.3009$  & $0.5419 \pm 0.2801$   & $0.5198 \pm 0.2870$ & $0.5197\pm0.3014$ & $0.4654\pm0.3006$ & $0.5121\pm0.2983$\\ \hline
  & \multicolumn{6}{c}{Random Location: $n_1 = 5, n_2 = 20, n_3 = 100$} \\\hline
   Percentage of Points in Key Shape & T-ANOVA & Heo's ANOVA & PERMANOVA & T-ANOVA & Heo's ANOVA & PERMANOVA\\\hline
  100\% & $0.4411 \pm 0.2541$ & $0.4738 \pm 0.3197$   & $0.5481 \pm 0.2707$ & $0.4958\pm0.2995$ & $0.4987\pm0.2799$ & $0.5257\pm0.3015$\\
  95\% & $0.4498 \pm 0.2774$ 	& $0.4623 \pm 0.2825$   & $0.5271 \pm 0.2884$ & $0.4935\pm0.3108$ & $0.4584\pm0.2931$ & $0.5271\pm0.2757$ \\
  90\% & $0.4670 \pm 0.2849$  & $0.4953 \pm 0.2868$   & $0.4747 \pm 0.2870$ & $0.4856\pm0.2607$ & $0.5109\pm0.2989$ & $0.5436\pm0.2767$ \\ \hline
  \end{tabular}}
  \caption{\label{tab: results_anova_robustness} Summary of mean$\pm$standard deviation of $p$-values of the T-ANOVA, Heo's ANOVA, and PERMANOVA in 100 simulations. Top half: In each simulation, the test was used to compare Group 1, 2, 3 of respective $n_1, n_2, n_3$ random samples were generated with a pre-specified percentage (90\%, 95\%, 100\%) of 100 or 200 points from the original key shape "contaminated" with topological noise in the form of a much smaller 1-cycle next to the keyhole with pre-specified (in such case $n_1 = n_2 = n_3 = 4$) or random locations (in such case $n_1 = n_2 = n_3 = 5$, $n_1 = n_2 = n_3 = 20$, or $n_1 = 5, n_2 = 20, n_3 = 100$). Bottom half: In each simulation, the test was used to compare Group 1, 2, 3 of respective $n_1, n_2, n_3$ random samples were generated with a pre-specified percentage (90\%, 95\%, 100\%) of 100 or 200 points from the original key shape with a pre-specified (in such case $n_1 = n_2 = n_3 = 4$) or random (in such case $n_1 = n_2 = n_3 = 5$, $n_1 = n_2 = n_3 = 20$, or $n_1 = 5, n_2 = 20, n_3 = 100$) quarter of the keyhole left.}
  \end{table}

\noindent{\em Simulation results under different bandwidths.} Tables~\ref{tab: results_pt_bd5} and \ref{tab: results_net_bd5} summarize the results of simulation studies under different bandwidths. 

\begin{table}[h!]
  \begin{tabular}{ccc}
  \hline
  \multicolumn{3}{c}{{\bf Simulation Studies based on Point Clouds}} \\\hline
  \multicolumn{3}{c}{\textbf{Performance of Two-sample Test}} \\\hline
  \multicolumn{3}{c}{\textbf{Sensitivity in Detecting Differential Hole Presence Among Multiple Groups}} \\\hline
  & $\sigma=5$ & $\sigma=10$\\\hline
  $m = 20, n = 20$ & $0.0000\pm0.0000$ & $0.0000\pm0.0000$ \\ \hline
  \multicolumn{3}{c}{\textbf{Robustness under Variation of Topological Noise}} \\\hline
    & $\sigma=5$ & $\sigma=10$\\\hline
  Pre-specified Location: $m = 4, n = 4$  & $0.4596 \pm 0.3093$ & $0.4795 \pm 0.3038$  \\
  Random Location: $m = 20, n = 20$  & $0.4849 \pm 0.2945$ & $0.4817 \pm 0.2951$ \\ \hline
  \multicolumn{3}{c}{\textbf{Robustness under Variation of Hole Location}} \\\hline
  & $\sigma=5$ & $\sigma=10$\\\hline
  Pre-specified Location: $m = 4, n = 4$ & $0.4443 \pm 0.2988$ & $0.4469 \pm 0.2960 $ \\
  Random Location: $m = 20, n = 20$  & $0.4862 \pm 0.2793$ & $0.4797 \pm 0.2813$ \\ \hline
  \multicolumn{3}{c}{\textbf{Performance of ANOVA}} \\\hline
  \multicolumn{3}{c}{\textbf{Sensitivity in Detecting Differential Hole Presence Among Multiple Groups}} \\\hline
  & $\sigma=5$ & $\sigma=10$\\\hline
  $n_1 = n_2 = n_3 = 20$ & $0.0000\pm0.0000$ & $0.0000\pm0.0000$ \\ \hline
  \multicolumn{3}{c}{\textbf{Robustness under Variation of Topological Noise}} \\\hline
    & $\sigma=5$ & $\sigma=10$ \\\hline
  Pre-specified Location: $n_1 = n_2 = n_3 = 4$ & $0.4802 \pm 0.3102$ & $0.4845\pm0.3094$  \\
  Random Location: $n_1 = n_2 = n_3 = 20$ & $0.4655 \pm 0.2983$ & $0.4649\pm0.3022$ \\ \hline
  \multicolumn{3}{c}{\textbf{Robustness under Variation of Hole Location}} \\\hline
  & $\sigma=5$ & $\sigma=10$\\\hline
  Pre-specified Location: $n_1 = n_2 = n_3 = 4$ & $0.4887 \pm 0.3121$ & $0.4857 \pm 0.3033$  \\
  Random Location: $n_1 = n_2 = n_3 = 20$ & $0.4924 \pm 0.2776$ & $0.4869\pm 0.2726$ \\ \hline
  \end{tabular}
  \caption{\label{tab: results_pt_bd5} Summary of mean$\pm$standard deviation of $p$-values of the T-ANOVA with bandwidth $\sigma=5$ and $\sigma=10$ in 100 simulations. In each simulation, the test is used to compare Group 1, 2, 3 of respective $n_1, n_2, n_3$ random samples are generated with a pre-specified percentage $95\%$ of 200 points.}
  \end{table}

\begin{table}[h!]
  \begin{tabular}{ccc|ccc}
  \hline
  \multicolumn{6}{c}{{\bf Simulation Studies based on Real Brain Networks}} \\\hline
 \multicolumn{6}{c}{\textbf{Robustness under Slight Perturbations to Network Topology}} \\\hline
  \multicolumn{3}{c}{\textbf{Performance of Two-sample Tests}} & \multicolumn{3}{c}{\textbf{Performance of ANOVA}}\\\hline
  & $\sigma=5$ & $\sigma=10$ & & $\sigma=5$ & $\sigma=10$ \\\hline
  $20~\text{VS}~20$ & $1.0000\pm0.0000$ & $1.0000\pm0.0000$ & $20~\text{VS}~20~\text{VS}~20$ & $1.0000\pm0.0000$ & $1.0000\pm0.0000$ \\ \hline
  \multicolumn{6}{c}{\textbf{Sensitivity in Detecting Differential Lesion Presence}} \\\hline
  \multicolumn{3}{c}{\textbf{Performance of Two-sample Tests}} & \multicolumn{3}{c}{\textbf{Performance of ANOVA}}\\\hline
  \multicolumn{6}{c}{No Lesion Vs No Lesion Vs Frontal (4) \& Temporoparietal (4) \& Premotor (4)} \\\hline
  & $\sigma=5$ & $\sigma=10$ & & $\sigma=5$ & $\sigma=10$ \\\hline
  $20~\text{VS}~20$ & $0.0000\pm0.0000$ & $0.0000\pm0.0000$ & $20~\text{VS}~20~\text{VS}~20$ & $0.0006\pm0.0011$ & $0.0002\pm0.0007$\\\hline
  \multicolumn{6}{c}{No Lesion Vs No Lesion Vs Frontal (4) \& Temporoparietal (4) \& Premotor (4) \& Temporal(4)} \\\hline
  & $\sigma=5$ & $\sigma=10$ & & $\sigma=5$ & $\sigma=10$ \\\hline
  $20~\text{VS}~20$ & $0.0000\pm0.0000$ & $0.0000\pm0.0000$ & $20~\text{VS}~20~\text{VS}~20$ & $0.0000\pm0.0000$ & $0.0000\pm0.0000$ \\\hline
  \end{tabular}
  \caption{\label{tab: results_net_bd5} Summary of mean $\pm$ standard deviation of $p$-values of T-ANOVA in 100 simulations.}
  \end{table}


\bibliographystylesupp{plainnat}
\bibliographysupp{bib_top_inf_aoas} 

\end{document}